\documentclass[a4paper,twoside]{article}

\usepackage{times}

\usepackage{multicol}
\usepackage[bookmarks=true]{hyperref}
\hypersetup{
	colorlinks,
	linkcolor={red!50!black},
	citecolor={blue!50!black},
	urlcolor={blue!80!black}
}

\usepackage[dvipsnames,table]{xcolor}
\usepackage{amsmath,amssymb,amsfonts}
\usepackage{tikz, tikz-3dplot}
\usepackage{booktabs}
\usepackage{subfig}
\usepackage[binary-units=true]{siunitx}
\usepackage{multirow}
\usepackage{rotating} 
\usepackage{tabularx}
\usepackage{graphbox}
\usepackage[outline]{contour}
\usepackage{bm} 
\usepackage[nolist,nohyperlinks]{acronym}
\usepackage{pifont}
\usepackage{colortbl}     
\usepackage{todonotes}   
\usepackage{datetime} 
\usepackage{placeins} 

\usepackage{hhline} 

\usepackage{listings}
\definecolor{solarized-green}{RGB}{133,153,0}
\definecolor{solarized-blue}{RGB}{38,139,210}
\definecolor{solarized-magenta}{RGB}{211,54,130}
\lstdefinestyle{easypbr}{
	language=c,
	morekeywords = {apply, List},
	deletekeywords={try},
	otherkeywords = {;, |>, =>, >>, <+},
	keywordstyle=\bfseries\color{black!75}, 
	morekeywords= [2]{ 
		Mesh,
		Scene, 
		Viewer, 
		Recorder
	},
	keywordstyle= [2]{\bfseries\color{ph-orange}},
	morekeywords = [3]{ 
		V,NV,NF,E,F,L
	},
	keywordstyle= [3]{\bfseries\color{MidnightBlue}},
	morekeywords = [4]{ 
		show,
		record,
		load_environment_map,
		rotate_axis_angle,
		orbit,
		translate,
		update
	},
	keywordstyle = [4]{\bfseries\color{ph-blue}},
	morekeywords = [5]{
		bf,
		mm,
		threemaps,
		mt
	},
	keywordstyle = [5]{\bfseries\color{RedViolet}},
	morekeywords= [6]{ 
		|>, >>,
		fun,
		app,
		map,
		zip,
		reduce,reduceSeq,reduceSeqUnroll,
		mapSeq,mapSeqUnroll,
		mapPar,
		mapVec,
		asVector,
		asScalar,
		map2D,
		pad,
		pad2D,
		slide,
		slide2D,
		join,
		transpose,
		split
	},
	keywordstyle= [6]{\bfseries\color{OliveGreen}},
	morekeywords = [7]{ 
		dot
	},
	keywordstyle = [7]{\color{RawSienna}},
	morekeywords = [8]{ 
		weights2d,weightsH,weightsV,
		nbh,
		x,y,
		f,nf,
		g,
		xs,
		img,
		a, na,arow,bcol,
		b, nb
	},
	keywordstyle = [8]{\color{RedOrange}},
	basicstyle=\bfseries\ttfamily\footnotesize, 
	commentstyle=\itshape, 
	stringstyle=\itshape \color{ph-green}, 
	numbers=none, 
	tabsize = 2,
	xleftmargin=.0\parindent,
	numbersep=8pt, 
	showstringspaces=false, 
	breaklines=false, 
	aboveskip=.5em,
	belowskip=.5em,
	escapechar=@,
	captionpos=b,
	mathescape=true,
	literate={`}{\lq}1 
}

\usepackage{epsfig}
\usepackage{calc}
\usepackage{amssymb}
\usepackage{amstext}
\usepackage{amsmath}
\usepackage{amsthm}
\usepackage{multicol}
\usepackage{pslatex}
\usepackage{apalike} 
\usepackage{SCITEPRESS}  

\usepackage[outdir=./]{epstopdf}

\newcommand{\PreserveBackslash}[1]{\let\temp=\\#1\let\\=\temp}
\newcolumntype{C}[1]{>{\PreserveBackslash\centering}p{#1}}
\newcolumntype{R}[1]{>{\PreserveBackslash\raggedleft}p{#1}}
\newcolumntype{L}[1]{>{\PreserveBackslash\raggedright}p{#1}}

\newacro{TSDF}{truncated signed distance function}
\newacro{GUI}{graphical user interface}
\newacro{BRDF}{bidirectional reflectance distribution function}
\newacro{IBL}{image-based lighting}
\newacro{SSAO}{screen-space ambient occlusion}
\newacro{EDL}{eye dome lighting}
\newacro{SLAM}{simultaneous localization and mapping}
\newacro{PBR}{physically-based rendering}
\newacro{HDR}{high dynamic range}
\newacro{LDR}{low dynamic range}
\newacro{ACES}{Academy Color Encoding System}

\usetikzlibrary{calc,arrows,arrows.meta,backgrounds,tikzmark,shapes.geometric, decorations.pathreplacing,decorations.markings, spy}

\definecolor{ph-purple}{RGB}{129, 39, 232}
\definecolor{ph-blue}{RGB}{5, 131, 227}
\definecolor{ph-gray}{rgb}{0.5, 0.5, 0.5}
\definecolor{ph-orange}{RGB}{227, 127, 5}
\definecolor{ph-green}{RGB}{0, 135, 124}
\definecolor{ph-yellow}{RGB}{235, 201, 52}
\definecolor{ph-light-green}{RGB}{181, 209, 21}
\definecolor{ph-red}{RGB}{250, 101, 60}
\colorlet{ph-orange-light}{ph-orange!70}
\colorlet{ph-blue-light}{ph-blue!70}
\colorlet{ph-purple-light}{ph-purple!70}
\colorlet{ph-green-light}{ph-green!70}
\definecolor{ph-light-gray}{rgb}{0.75, 0.75, 0.75}

\newcommand{\reffig}[1]{Fig.~\ref{#1}}
\newcommand{\reftab}[1]{Tab.~\ref{#1}}

\begin{document}
	

\newcommand\blfootnote[1]{%
	\begingroup
	\renewcommand\thefootnote{}\footnote{#1}%
	\addtocounter{footnote}{-1}%
	\endgroup
}

\title{EasyPBR: A Lightweight Physically-Based Renderer}

\author{
	\authorname{Radu Alexandru Rosu\orcidAuthor{0000-0001-7349-4126} and Sven Behnke\orcidAuthor{0000-0002-5040-7525}}
	\affiliation{ 
		Autonomous Intelligent Systems, University of Bonn, Germany
	}
	\email{ rosu@ais.uni-bonn.de, behnke@ais.uni-bonn.de}
}

\keywords{Physically-Based Rendering, Synthetic Data Generation, Visualization Toolkit}

\abstract{
	Modern rendering libraries provide unprecedented realism, producing real-time photorealistic 3D graphics on commodity hardware. Visual fidelity, however, comes at the cost of increased complexity and difficulty of usage, with many rendering parameters requiring a deep understanding of the pipeline.
	We propose EasyPBR as an alternative rendering library that strikes a balance between ease-of-use and visual quality. EasyPBR consists of a deferred renderer that implements recent state-of-the-art approaches in physically based rendering. It offers an easy-to-use Python and C++ interface that allows high-quality images to be created in only a few lines of code or directly through a graphical user interface. The user can choose between fully controlling the rendering pipeline or letting EasyPBR automatically infer the best parameters based on the current scene composition.  
	The EasyPBR library can help the community to more easily leverage the power of current GPUs to create realistic images. These can then be used as synthetic data for deep learning or for creating animations for academic purposes.
}

\global\csname @topnum\endcsname 0
\global\csname @botnum\endcsname 0

\onecolumn \maketitle \normalsize \setcounter{footnote}{0} \vfill

\section{Introduction}

Modern rendering techniques have become advanced enough for photorealistic images to be produced even on commodity hardware. Advances such as real-time ray tracing and physically-based materials have allowed current rendering pipelines to closely follow the theoretical understanding of light propagation and how it interacts with the real world.
However, such advancements in rendering come with an increase in complexity for the end user, often requiring a deep understanding of the rendering pipeline to achieve good results. \blfootnote{This work has been funded by the Deutsche Forschungsgemeinschaft (DFG, German Research Foundation) under Germany's Excellence Strategy - EXC 2070 - 390732324 and by the German Federal Ministry of Education and Research (BMBF) in the project ”Kompetenzzentrum: Aufbau des Deutschen Rettungsrobotik-Zentrums” (A-DRZ).}

Our proposed EasyPBR addresses this issue by offering a 3D viewer for visualizing various types of data (meshes, point clouds, surfels, etc.) with high-quality renderings while maintaining a low barrier of entry. Scene setup and object manipulation can be done either through Python or C++. Furthermore, meshes can be manipulated through the powerful libigl~\cite{libigl} library for geometry processing since our mesh representation shares a common interface. The user can choose to configure rendering parameters before the scene setup or at runtime through the GUI. If the parameters are left untouched, EasyPBR will try to infer them in order to best render the given scene. EasyPBR uses state-of-the-art rendering techniques and offers easy extensions for implementing novel methods through using a thin abstraction layer on top of OpenGL.

EasyPBR and all the code needed to reproduce the figures in this paper is made available at \\ 
 \url{https://github.com/AIS-Bonn/easy_pbr} \\
A  video  with additional footage is also available online~\footnote{\url{http://www.ais.uni-bonn.de/videos/GRAPP_2020_Rosu/}}.

\noindent Our main contributions are:  
\begin{itemize}
	\item a lightweight framework for real-time physically-based rendering,
	\item an easy-to-use Python front-end for scene setup and manipulation, and
	\item powerful mesh manipulation tools through the libigl~\cite{libigl} library.
\end{itemize}

\section{Related Work} 

Various 3D libraries currently offer rendering of high fidelity visuals. Here we compare against the most widely used ones.

Meshlab~\cite{cignoni2008meshlab} is a popular open-source tool for processing, editing, and visualizing triangular meshes. Its functionality can be accessed either through the \ac{GUI} or the provided scripting interface. This makes Meshlab difficult to integrate into current Python or C++ projects. In contrast, EasyPBR offers both a Python package that can be easily imported and a shared library that can be linked into an existing C++ project. EasyPBR also integrates with libigl~\cite{libigl}, allowing the user to access powerful tools for geometry processing. Additionally, EasyPBR offers more realistic renderings of meshes together with functionality for creating high-resolution screenshots or videos.

Blender~\cite{blender} is an open-source 3D creation suite. It includes all aspects of 3D creation, from modeling to rendering and video editing; and it offers a Python API, which can be used for scripting. However, the main usage of Blender is to create high-quality visuals through ray-traced rendering, which is far from real-time capable. The Python API is also not the main intended use case of Blender, and while rendering commands can be issued through scripts, there is no visual feedback during the process.
In contrast, we offer real-time rendering and control over the scene from small Python or C++ scripts.
 
VTK~\cite{vtk} is an open-source scientific analysis and visualization tool. While initially its main rendering method was based on Phong shading, recently a physically-based renderer together with \ac{IBL} has also been included. Extensions of the main rendering model with new techniques is cumbersome as it requires extensive knowledge of the VTK framework. 
In contrast, our rendering methods are easy to use and we keep a thin layer of abstraction on top of OpenGL for simple extendibility using custom callbacks.
 
Marmoset toolbag~\cite{marmoset} is a visual tool designed to showcase 3D art. It features a real-time PBR renderer, which allows easy setup of a scene to create high-quality 3D presentations. However, it is not available on Linux and is also distributed under a paid license.  

Unreal Engine~\cite{unreal} is a state-of-the-art engine created with the goal to provide real-time high-fidelity visuals. It has been used in professional game-making, architecture visualization, and VR experiences. While it provides a plethora of tools for content creation, the entry barriers can also be quite high. Additionally, the Python API provided can only be used as an internal tool for scripting and results in cumbersome setup code for even easy importing of assets and rendering. In contrast, EasyPBR acts as a Python library that can be readily imported in any existing project and used to draw to screen in only a couple of lines of code.

We showcase results from EasyPBR compared with VTK and Marmoset in~\reffig{fig:heads}. We use the high-quality 3D scan from~\cite{Scanstore} and render it under similar setups. While our renderer does not feature sub-surface scattering shaders like Marmoset, it can still achieve high-quality results in less than 10 lines of code. In contrast, for similar results, VTK requires more than 150 lines in which the user needs to manually define rendering passes for effects such as shadows baking and shadow mapping.

\bgroup
\def\ImgHead{./imgs/head/vtk_crop_sized.jpg} 
\newlength{\WHead}
\newlength{\HHead}
\settowidth{\WHead}{\includegraphics{\ImgHead}}
\settoheight{\HHead}{\includegraphics{\ImgHead}}
\begin{figure}[]
	\captionsetup[subfloat]{labelformat=empty}
	\centering
	
	\subfloat
	[]
	[ \centering 
	a) Marmoset 
	
	\scriptsize \cite{marmoset}  ] 
	{
		\includegraphics[trim=0.28\WHead{} 0.28\HHead{} 0.52\WHead{} 0.31\HHead{},clip,width=0.3\columnwidth] {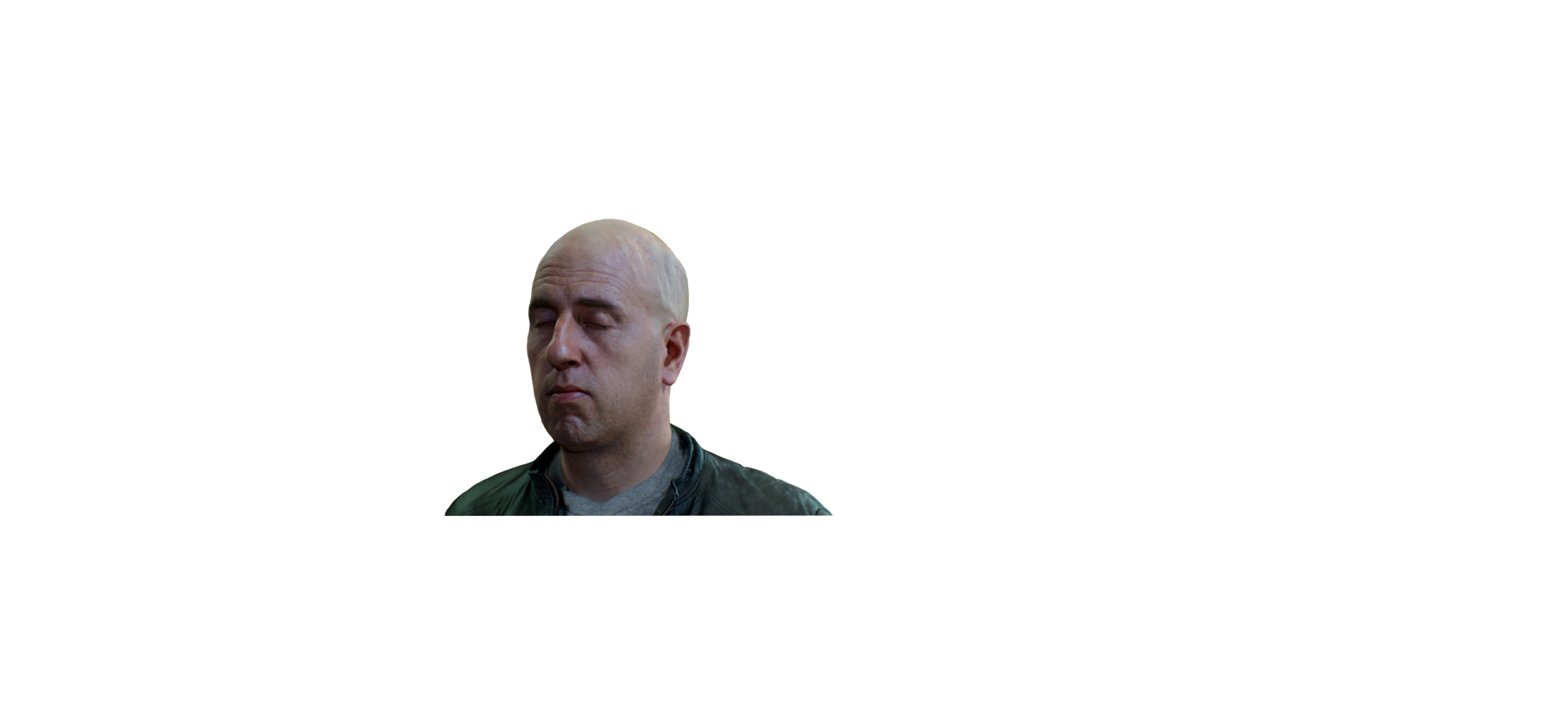}
		\label{fig:head1}
	}
	\subfloat
	[]
	[ \centering 
	b) EasyPBR 
	
	(Ours)  ] 
	{
		\includegraphics[trim=0.28\WHead{} 0.28\HHead{} 0.52\WHead{} 0.31\HHead{},clip,width=0.3\columnwidth] {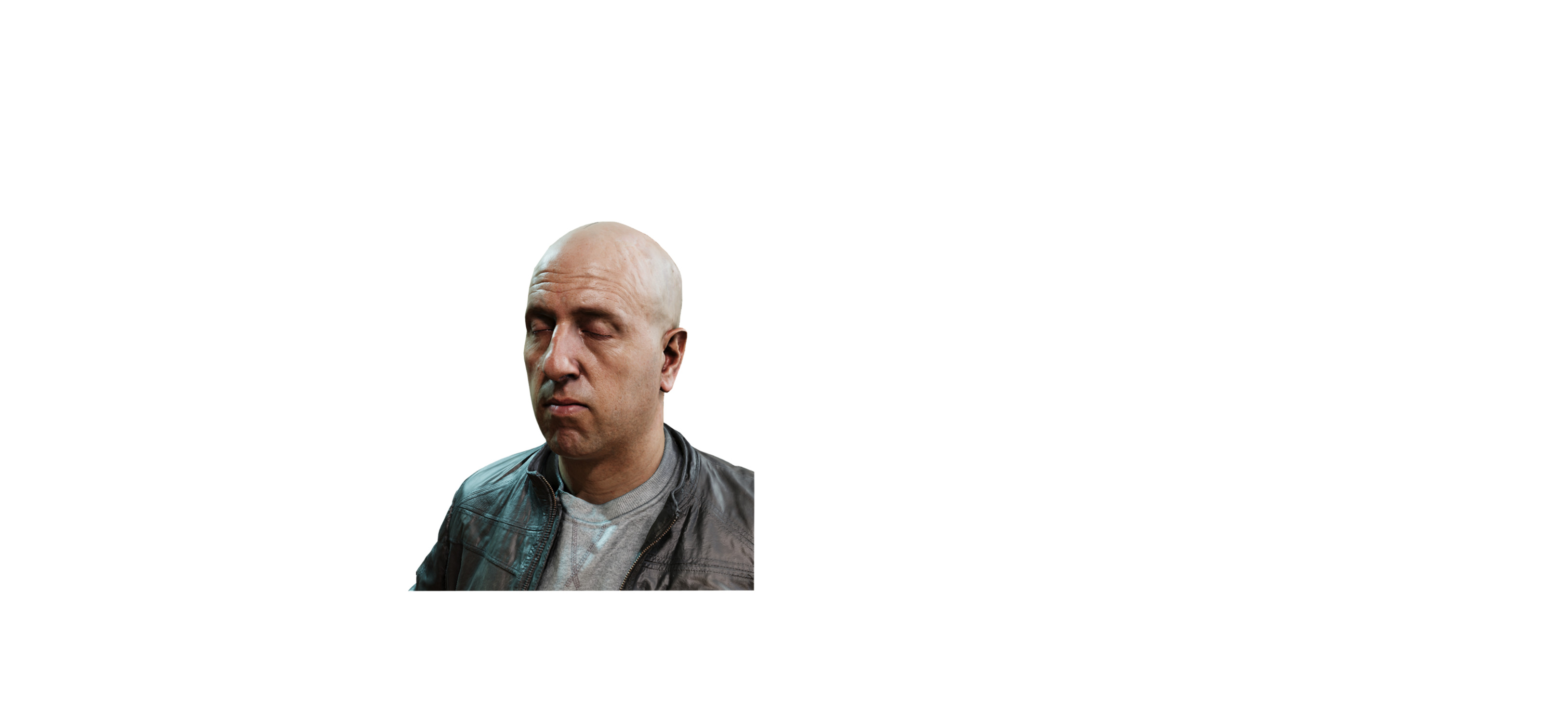}
		\label{fig:head2}
	}
	\subfloat
	[]
	[ \centering 
	c) VTK 
	
	\scriptsize \cite{vtk}  ] 
	{
		\includegraphics[trim=0.28\WHead{} 0.28\HHead{} 0.52\WHead{} 0.31\HHead{},clip,width=0.3\columnwidth] {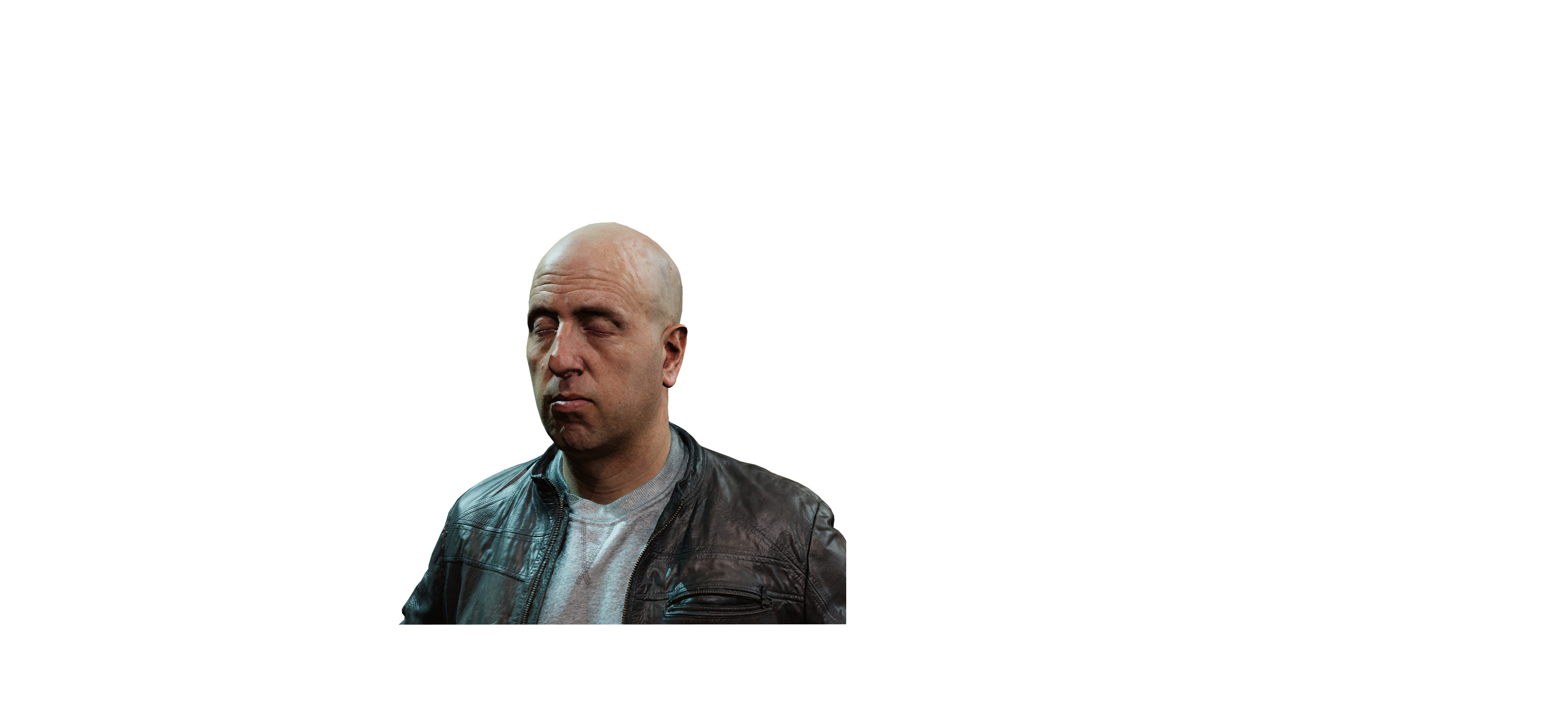}
		\label{fig:head3}
	}
	
	\caption{ Comparison of various PBR rendering tools. }
	\label{fig:heads}
\end{figure}
\egroup

\section{Physically-based Rendering }
	Physically-based rendering or PBR is a set of shading models designed to achieve high realism though accurately modeling light and material interaction. Previous shading models like Phong shading are not based on mathematically rigorous analysis of light and can lead to unrealistic results. PBR attempts to address this issue by basing the shading equations on the laws of light interaction.

	PBR follows the mathematical modeling of light based on the reflectance equation:
	\begin{equation}
		L_o(p,\omega_o) = \int_{\Omega} f_r(p, \omega_i, \omega_o) L_i(p,\omega_i) (n \cdot \omega_i) d\omega_i,
	\end{equation} 
	where $L_o(p,\omega_o)$ is the outgoing radiance from point $p$ in direction $\omega_o$ which gathers over the hemisphere $\Omega$ the incoming radiance $L_i$ weighted by the BRDF $f_r(p, \omega_i, \omega_o)$ and the angle of incidence between the incoming ray $\omega_i$ and the surface normal $n$.
	
	To model materials in a PBR framework we use the Cook-Torrance~\cite{cook1982reflectance} BRDF. Material properties are specified by two main parameters: metalness and roughness. These parameters cover the vast majority of the real-world materials. By using a physically-based renderer, they will look realistic under different illumination conditions.
	
	Since Cook-Torrance is just an approximation of the underlying physics and there are many variants used in literature, some more realistic, others more efficient, we choose the same approximation used in Unreal Engine 4~\cite{karis2013real} which strikes a good balance between realism and efficiency.

	\subsection{Image-based Lighting}

	 To fully solve the reflectance equation, light incoming onto the surface would have to be integrated over the whole hemisphere. However, this integral is not tractable in practice, and therefore, one approximation would be to gather only the direct contributions of the light sources in the scene. This has the undesirable effect of neglecting secondary bounces of light and causing shadows to be overly dark, yielding a non-realistic appearance. 
	 To address this, we use image-based lighting, which consists of embedding our 3D scene inside a \ac{HDR} environment cubemap in which every pixel acts as a source of light. This greatly enhances the realism of the scene and gives a sense that our 3D models "belong" in a certain environment as changes in the \ac{HDR} cubemap have a visible effect on the model's lighting. 
	
	Efficient sampling of the radiance from the environment map is done through precomputing increasingly blurrier versions of the cubemap, allowing for efficient sampling at runtime of only one texel that corresponds with a radiance over a large region of the environment map. 
	Specular reflections are also precomputed using the split-sum approximation. For more detail, we refer to the excellent article from Epic Games~\cite{karis2013real}. 
	
	We further extend the \ac{IBL} by implementing the approach of~\cite{FdezAguera2019Scattering} which further improves the visual quality of materials by taking into account multiple scatterings of light with only a slight overhead in performance.

\section{Deferred Rendering} 
	Rendering methods are often divided in two groups: forward rendering and deferred rendering, both with different pros and cons.
	
	Forward rendering works by rendering the whole scene in one pass, projecting every triangle to the screen and shading in one render call. This has the advantage of being simple to implement but may suffer from overdraw as having a lot of overlapping geometry causes much wasted effort in shading and lighting.
	
	Deferred rendering attempts to solve this issue by delaying the shading of the scene to a second step. The first step of a deferred renderer writes the material properties of the scene into a screen-size buffer called the G-Buffer. The G-Buffer typically records the position of the fragments, color, and normal. A second rendering pass reads the information from the G-Buffer and performs the light calculations. This has the advantage of performing costly shading operations only for the pixels that will actually be visible in the final image. 
	
	EasyPBR uses deferred rendering as its performance scales well with an increasing number of lights. Additionally, various post-processing effects like \ac{SSAO} are easier to implement in a deferred renderer than a forward one since all the screen-space information is already available in the G-Buffer.
	
	\bgroup
	\def\Img{./imgs/gbuffer/multichannel_18_outline2.jpg} 
	\newlength{\WSG}
	\newlength{\HSG}
	\settowidth{\WSG}{\includegraphics{\Img}}
	\settoheight{\HSG}{\includegraphics{\Img}}
	\def\Angle{-42.5}
	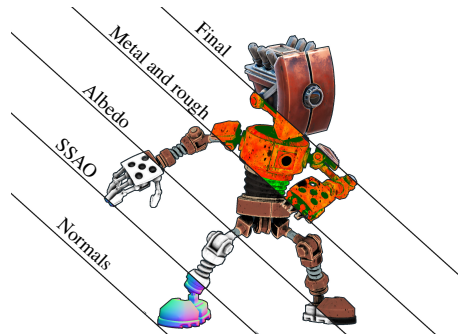
\begin{figure}
		\centering
		\resizebox{0.8\columnwidth}{!}{
			\begin{tikzpicture}
			\node[anchor=south west,inner sep=0] at (0,0) {\includegraphics[trim=0.15\WSG{} 0.1\HSG{} 0.2\WSG{} 0.05\HSG{},clip,width=8cm]{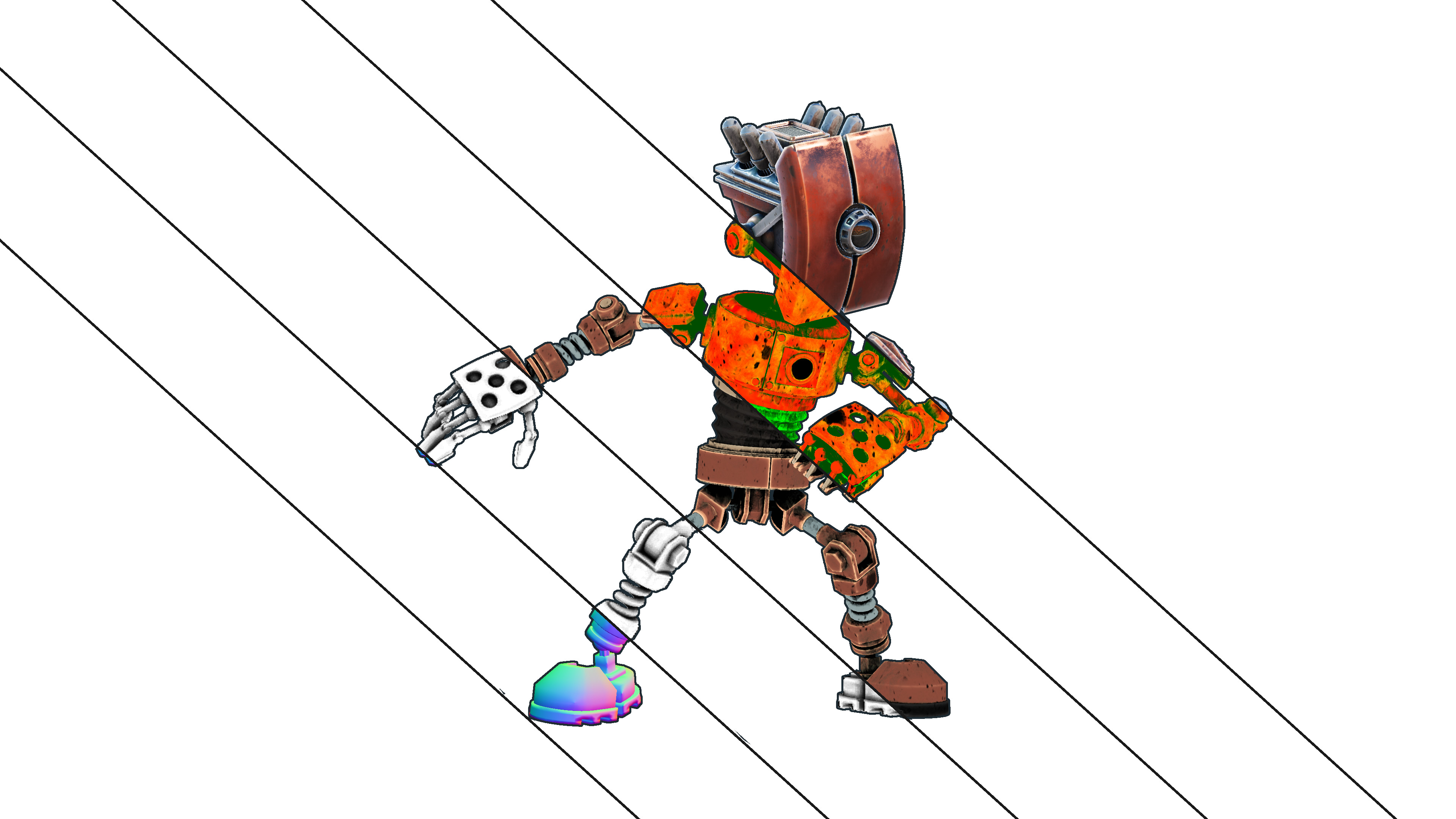}};

			\node [rotate=\Angle,anchor=north,outer sep=-4pt] at (3.7,5.4)  {Final};
			\node [rotate=\Angle,anchor=north,outer sep=-4pt] at (2.75,4.8)  {Metal and rough};
		
			\node [rotate=\Angle,anchor=north,outer sep=-4pt] at (1.8,4.2)  {Albedo};
			\node [rotate=\Angle,anchor=north,outer sep=-4pt] at (1.25,3.25)  {SSAO};
			\node [rotate=\Angle,anchor=north,outer sep=-4pt] at (1.4,1.70)  {Normals};
			\end{tikzpicture}
		} 
		\caption{The various G-Buffer channels together with Ambient Occlusion are composed into one final texture to be displayed on the screen. Here we display slices of each channel that is used for compositing.}
		\label{fig:gbufferChannels}
	\end{figure}
	\egroup

	\bgroup
	\newcolumntype{Q}{C{1.1cm}}
	\definecolor{g-green}{RGB}{49, 158, 78}
	\definecolor{g-red}{RGB}{204, 27, 24}
	\definecolor{g-blue}{RGB}{20, 27, 204}
	\definecolor{g-gray}{RGB}{100, 100, 100}
	\definecolor{t-gray}{RGB}{230, 230, 230} 
	\definecolor{t-gray-2}{RGB}{200, 200, 200}
	\definecolor{t-gray-3}{RGB}{120, 120, 120} 
	\definecolor{t-white}{RGB}{255, 255, 255}
	
	\renewcommand{\arraystretch}{1.2}
	
	\arrayrulecolor{t-white}
	\def\clinecolor{\hhline{|>{\arrayrulecolor{t-gray}}-%
			>{\arrayrulecolor{t-white}}|-|-|}}
	\begin{table}[ht]
		\caption{We structure the G-Buffer into four render targets.}
		\label{tab:gbufferLayout}
		
		\begin{center}
			\resizebox{1.0\columnwidth}{!}{
			\begin{tabular}{c|c|Q|Q|Q|Q}
			
				\multicolumn{1}{c|}{\multirow{2}{*}{ \cellcolor{t-gray-3} }} & \multicolumn{1}{c|}{ \cellcolor{t-gray-3} } & \multicolumn{4}{c}{\cellcolor{t-gray-3} \textcolor{t-white}{ \textbf{Usage} } } \\

				\hhline{>{\arrayrulecolor{t-gray-3}}-->{\arrayrulecolor{t-white}}----}

				\multicolumn{1}{c|}{ \cellcolor{t-gray-3} }& \multicolumn{1}{c|}{ \multirow{-2}{*}{ \cellcolor{t-gray-3} \textcolor{t-white}{ \textbf{Format} } } } & \cellcolor{g-red} \textcolor{t-white}{ \textbf{R} }&\cellcolor{g-green} \textcolor{t-white}{ \textbf{G} }&\cellcolor{g-blue} \textcolor{t-white}{ \textbf{B} }&\cellcolor{g-gray} \textcolor{t-white}{ \textbf{A} }\\

				\hline

				\cellcolor{t-gray}RT0 & \cellcolor{t-gray}RGBA8 & \multicolumn{3}{c|}{\cellcolor{t-gray}Albedo} & \multicolumn{1}{c}{\cellcolor{t-gray}Weight} \\
				\hline 
				\cellcolor{t-gray-2}RT1 & \cellcolor{t-gray-2}RGB8 & \multicolumn{3}{c|}{\cellcolor{t-gray-2}Normals} & \multicolumn{1}{c}{\cellcolor{t-gray-2}Unused}  \\
				\hline 
				\cellcolor{t-gray}RT2 & \cellcolor{t-gray}RG8 & \cellcolor{t-gray}Metal & \cellcolor{t-gray}Rough & \multicolumn{2}{c}{\cellcolor{t-gray}Unused} \\
				\hline 
				\cellcolor{t-gray-2}RT3 & \cellcolor{t-gray-2}R32 & \cellcolor{t-gray-2}Depth &  \multicolumn{3}{c}{\cellcolor{t-gray-2}Unused} \\

			\end{tabular}
			}
		\end{center}
	\end{table}
	\arrayrulecolor{black}
	\def\clinecolor{\hhline{|>{\arrayrulecolor{black }}-%
			>{\arrayrulecolor{black }}|-|-|}}
	\egroup

	The layout of our G-Buffer is described in~\reftab{tab:gbufferLayout}. 

	Please note that in our implementation, we do not store the position of each fragment but rather store only the depth map as a floating-point texture and reconstruct the position from the depth value. This saves us from storing three float values for the position, heavily reducing the memory bandwidth requirements for writing and reading into the G-Buffer.
	We additionally store a weight value in the alpha channel of the first texture. This will be useful later when we render surfels which splat and accumulate onto the screen with varying weights. 
	Several channels in the G-Buffer are purposely left empty so that they can be used for further rendering passes. 

	A visualization of the various rendering passes and the final composed image is shown in~\reffig{fig:gbufferChannels}.

	\subsection{Object Representation}
		We represent objects in our 3D scene as a series of matrices containing per-vertex information and possible connectivity to create lines and triangles. The following matrices can be populated: 
		\begin{itemize}
			\item $\mathbf{V} \in \mathbb{R}^{(n\times 3)}$  vertex positions,
			\item $\mathbf{N} \in \mathbb{R}^{(n\times 3)}$  per-vertex normals,
			\item $\mathbf{C} \in \mathbb{R}^{(n\times 3)}$  per-vertex colors,
			\item $\mathbf{T} \in \mathbb{R}^{(n\times 3)}$  per-vertex tangent vectors,
			\item $\mathbf{B} \in \mathbb{R}^{(n\times 1)}$  per-vertex bi-tangent vector length,
			\item $\mathbf{F} \in \mathbb{Z}^{(n\times 3)}$ triangle indices for mesh rendering,
			\item $\mathbf{E} \in \mathbb{Z}^{(n\times 2)}$ edge indices for line rendering.
		\end{itemize} 
		Note that for the bi-tangent vector, we store only the length, as the direction can be recovered through a cross product between the normal and the tangent.
		This saves significant memory bandwidth and is faster than storing the full vector.

	\subsection{Mesh Rendering}
		Mesh rendering follows the general deferred rendering pipeline. The viewer iterates through the meshes in the scene and writes their attributes into the G-Buffer. The attributes used depend on the selected visualization mode for the mesh (either solid color, per-vertex color, or texture). 
		
		When the G-Buffer pass is finished, we run a second pass which reads from the created buffer and creates any effect textures that might be needed (\ac{SSAO}, bloom, shadows, etc.).
		 
		A third and final pass is afterwards run which composes all the effect textures and the G-Buffer into the final image using PBR and \ac{IBL}.

	\subsection{Point Cloud Rendering}
	
		Point cloud rendering is similar to mesh rendering, i.e. the attributes of the point cloud are written into the G-Buffer.
		 
		The difference lies in the compositing phase where PBR and \ac{IBL} cannot be applied due to the lack of normal information. Instead, we rely on \ac{EDL}~\cite{boucheny2011eye}, which is a non-realistic rendering technique used to improve depth perception. The only information needed for \ac{EDL} is a depth map. \ac{EDL} works by looking at the depth of adjacent pixels in screen space and darkening the pixels which exhibit a sudden change of depth in their neighborhood. The bigger the local difference in depth values is, the darker the color is. The effect of \ac{EDL} can be seen in~\reffig{fig:EDL}. 
		
		Additionally, by sacrificing a bit more performance, the user can also enable \ac{SSAO} which further enhances the depth perception by darkening crevices in the model.

		\bgroup
		\def\Img{./imgs/cloud/cloud_edl_2.jpg} 
		\newlength{\WSC}
		\newlength{\HSC}
		\settowidth{\WSC}{\includegraphics{\Img}}
		\settoheight{\HSC}{\includegraphics{\Img}}
		\begin{figure}[]
			\captionsetup[subfloat]{labelformat=empty, justification=centering}
			\centering
			
			\subfloat[a) Plain cloud]{
				\includegraphics[trim=0.20\WSC{} 0.2\HSC{} 0.3\WSC{} 0.215\HSC{},clip,width=0.28\columnwidth] {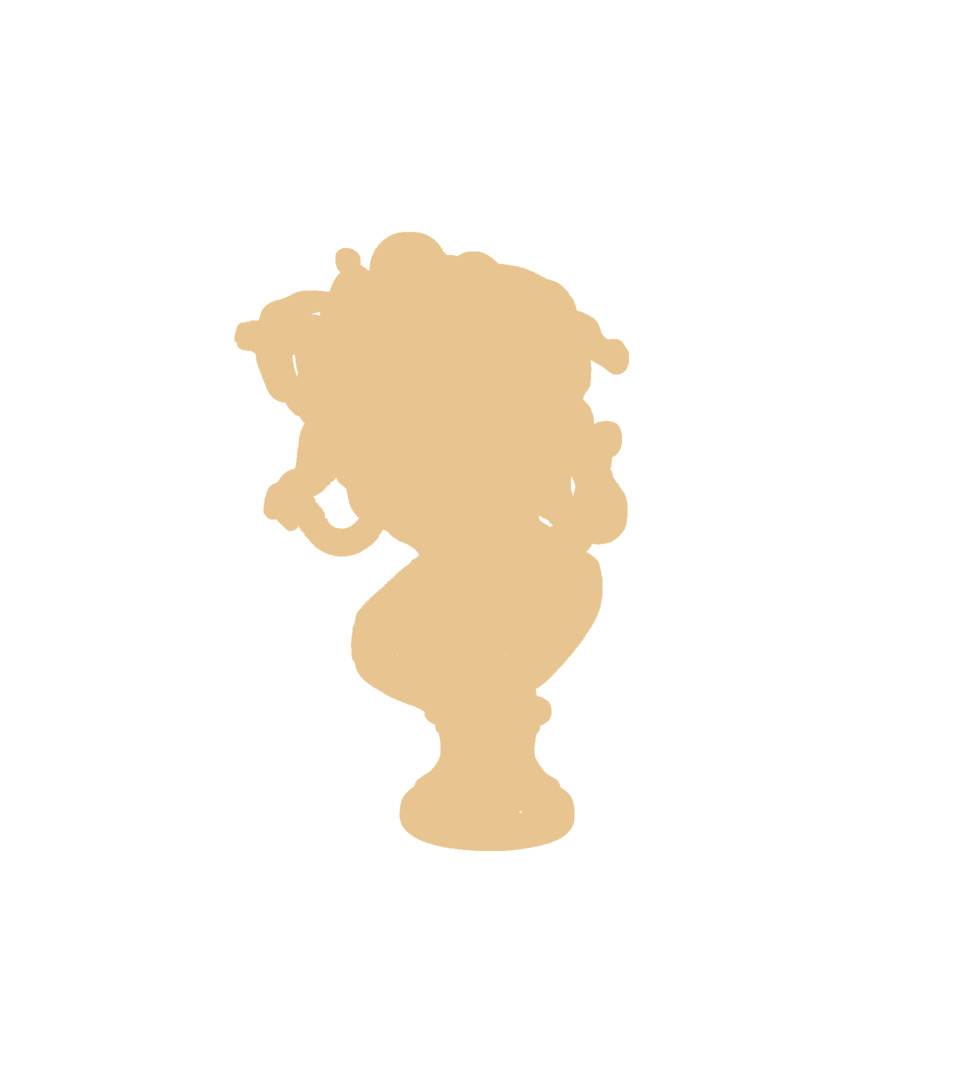}
			}
			\subfloat[b) EDL ]{
				\includegraphics[trim=0.20\WSC{} 0.2\HSC{} 0.3\WSC{} 0.215\HSC{},clip,width=0.28\columnwidth] {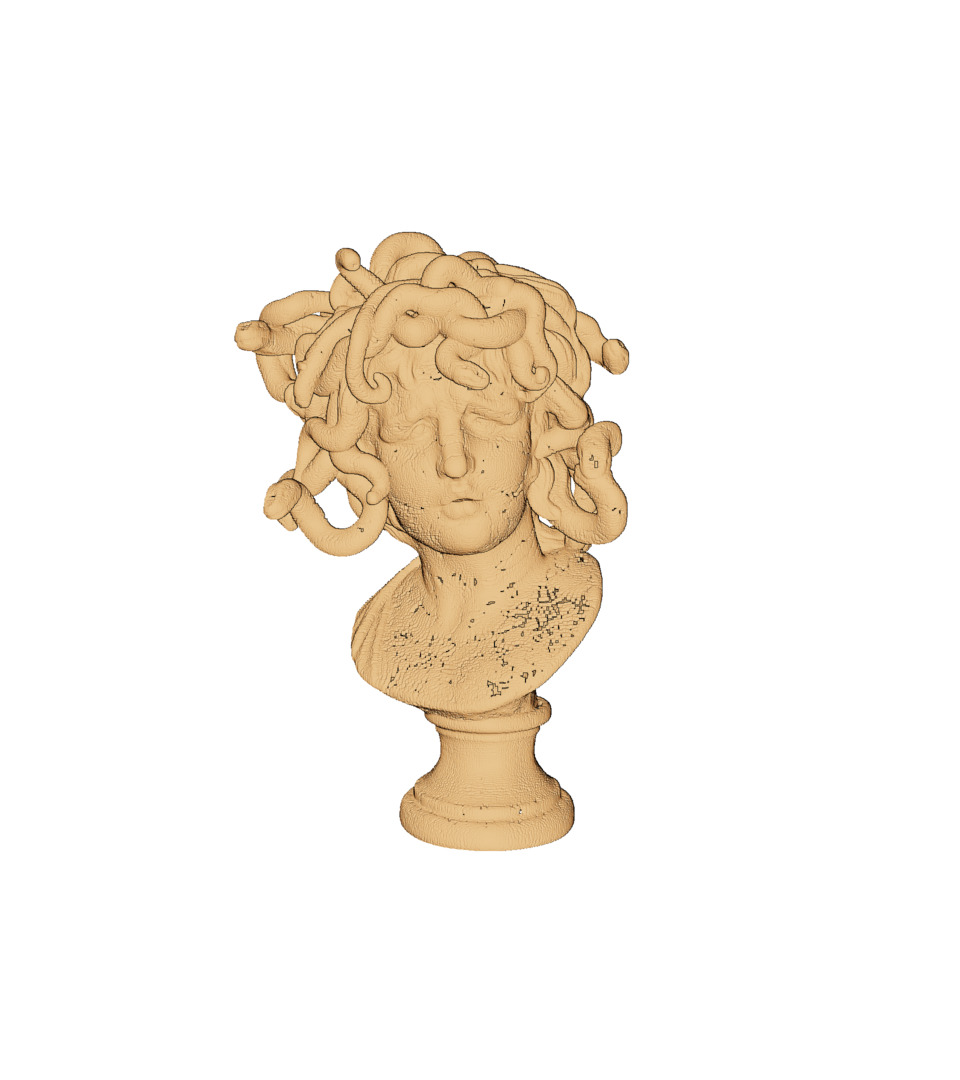}
			}
			\subfloat[c) EDL + SSAO]{
			\includegraphics[trim=0.20\WSC{} 0.2\HSC{} 0.3\WSC{} 0.215\HSC{},clip,width=0.28\columnwidth] {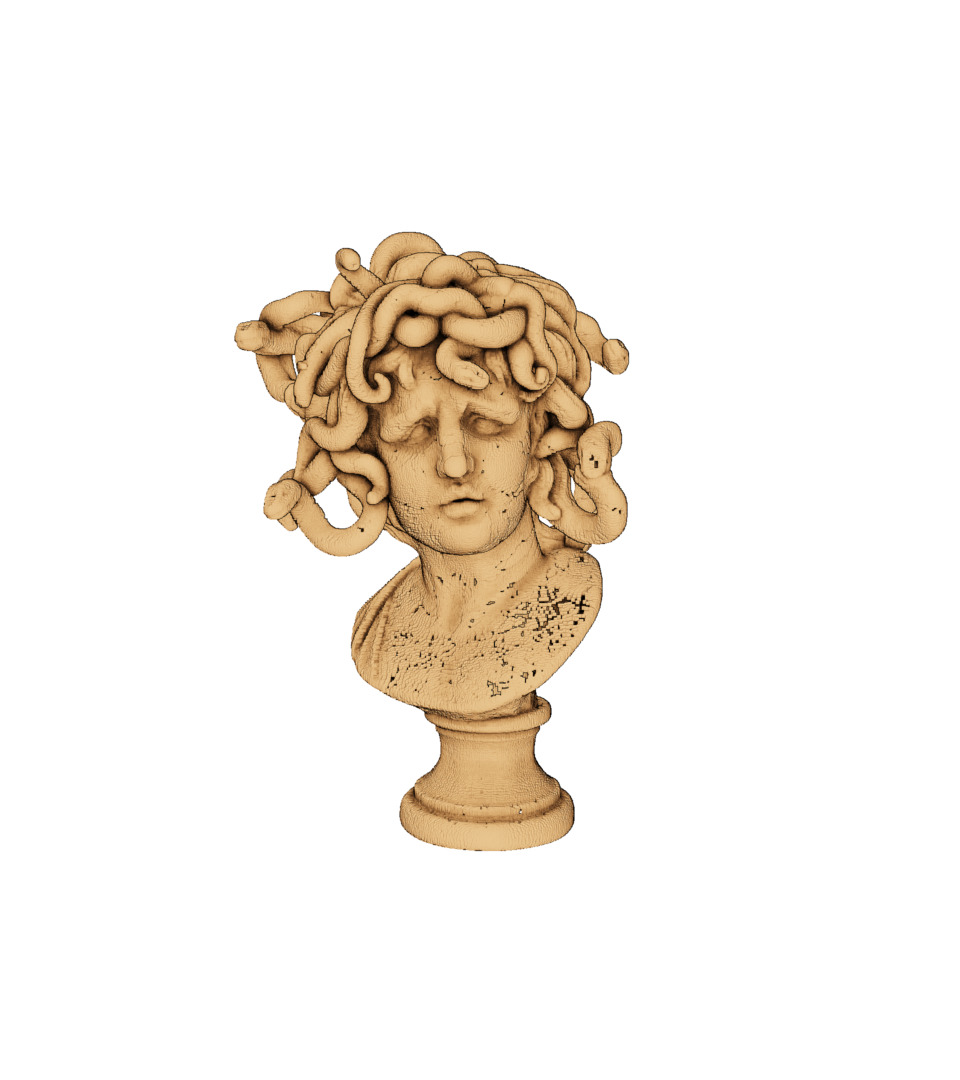}
			}
			
			
			\caption{ a) plain rendered point clouds results in flat shading and conveys little information. b) enabling Eye Dome Lighting gives a slight perception of depth, allowing the user to distinguish between various shapes. c) adding also Ambient Occlusion enhances the effect even further. }
			\label{fig:EDL}
		\end{figure}
		\egroup

	\subsection{Surfel Rendering}

	In various applications like \ac{SLAM} or 3D reconstruction, a common representation of the world is through surfels~\cite{droeschel2017continuous,stuckler2014multi}. Surfels are modeled as oriented disks with an ellipsoidal shape, and they can be used to model shapes that lack connectivity information. Rendering surfaces through surfels is done with splatting, which accumulates in screen space the contributions of various overlapping surfels. The three-step process of creating the surfels is ilustrated in~\reffig{fig:surfelCreate}.

		\bgroup
	\def\Img{./imgs/surfel/surfel_all.png} 
	\newlength{\WSSurf}
	\newlength{\HSSurf}
	\settowidth{\WSSurf}{\includegraphics{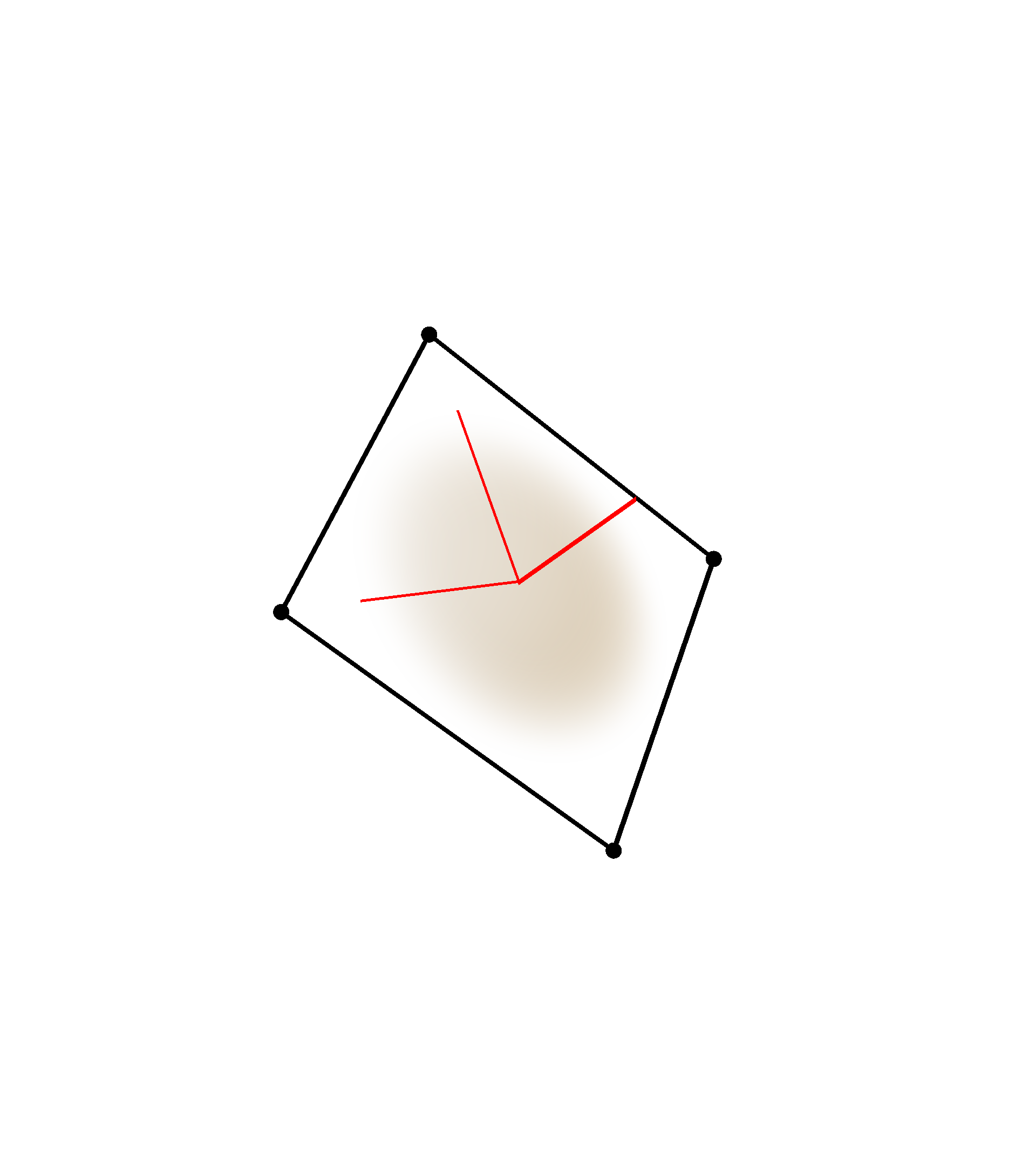}}
	\settoheight{\HSSurf}{\includegraphics{\Img}}
	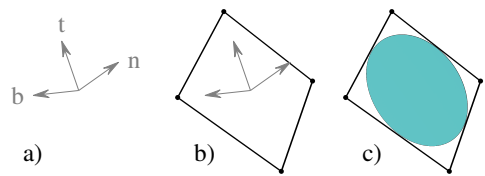
\begin{figure}[h]
		\captionsetup[subfloat]{labelformat=empty}
		\centering

		\resizebox{0.9\columnwidth}{!}{
			
			\begin{tikzpicture}[>={Stealth[inset=3pt,length=8pt,angle'=28,round]}]
			\node[anchor=south west,inner sep=0] at (0,0) {
				\includegraphics[trim=0.20\WSSurf{} 0.2\HSSurf{} 0.3\WSSurf{} 0.28\HSSurf{},clip,width=0.3\columnwidth] {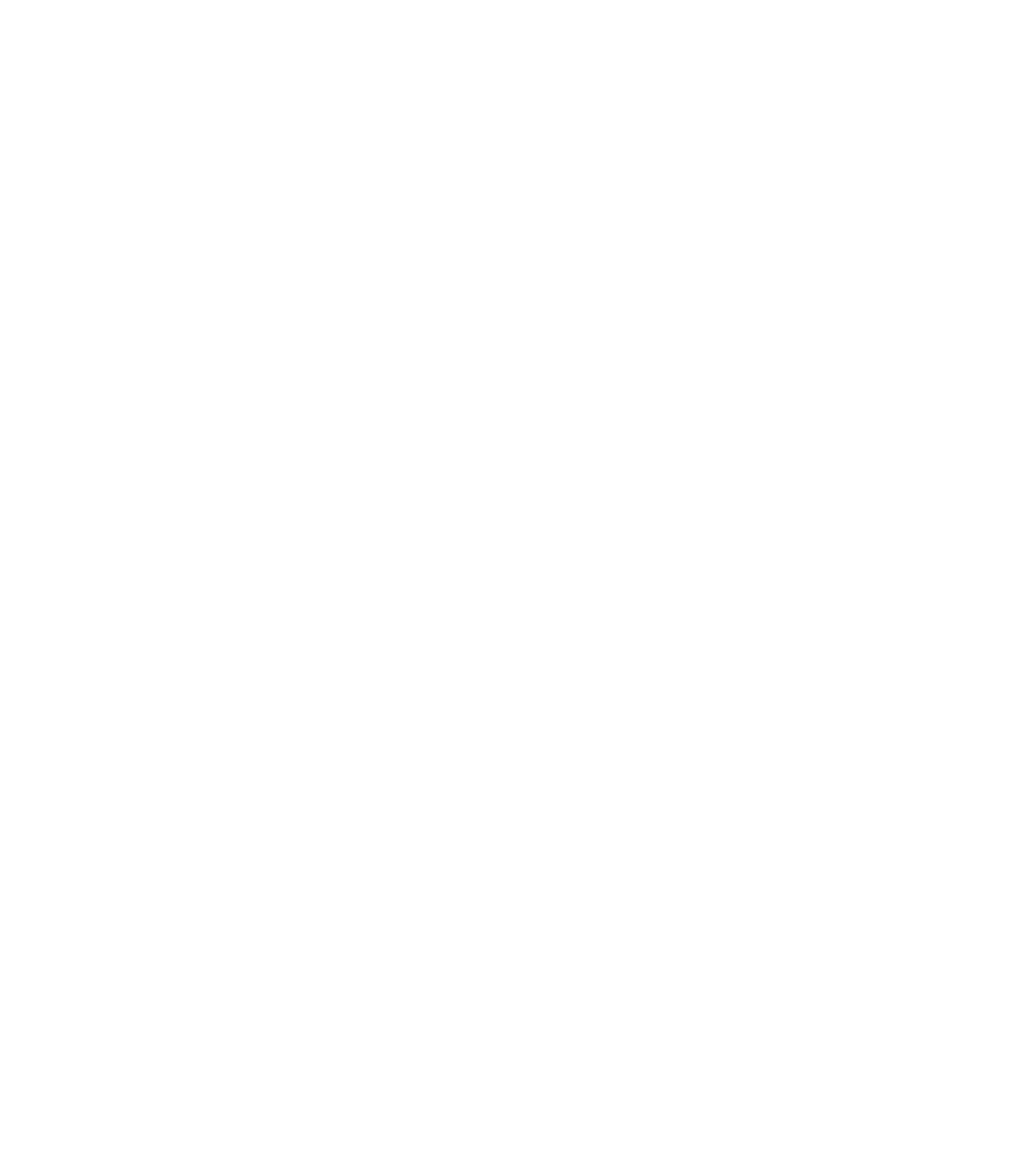}
			};

			\coordinate (origin) at (1.34, 1.5);
			\coordinate (n) at (1.9, 1.9);
			\coordinate (t) at (1.1, 2.2);
			\coordinate (b) at (0.7, 1.43);
			\draw[->,thin,gray] (origin)--(n) node[anchor=west]{n};
			\draw[->,thin,gray] (origin)--(t) node[anchor=south]{t};
			\draw[->,thin,gray] (origin)--(b) node[anchor=east]{b};
			
			\node[] at (0.7, 0.6) (label) {a)};
			\end{tikzpicture}
			\begin{tikzpicture}[>={Stealth[inset=3pt,length=8pt,angle'=28,round]}]
			\node[anchor=south west,inner sep=0] at (0,0) {
				\includegraphics[trim=0.20\WSSurf{} 0.2\HSSurf{} 0.3\WSSurf{} 0.28\HSSurf{},clip,width=0.3\columnwidth] {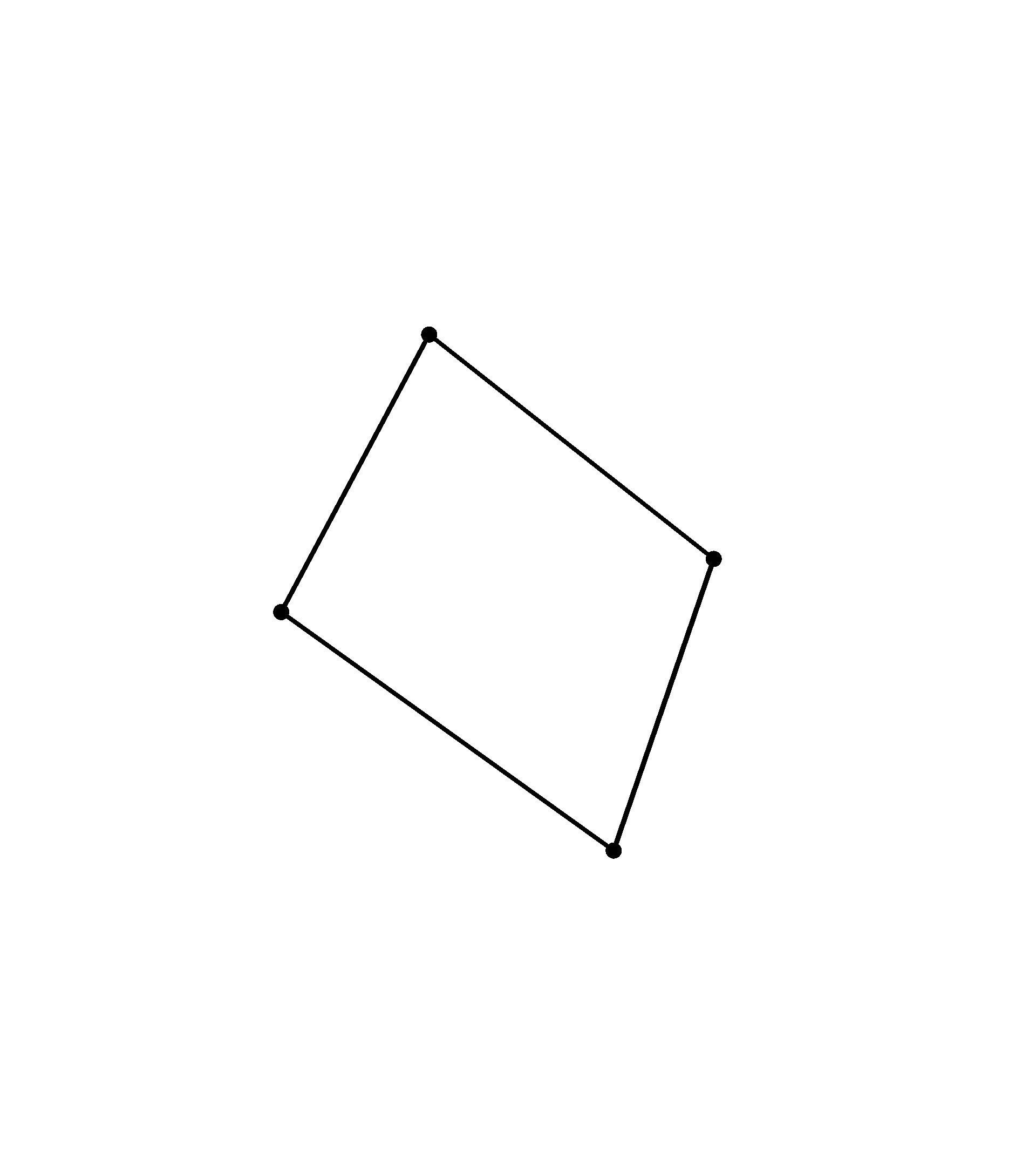}
			};
			\coordinate (origin) at (1.34, 1.5);
			\coordinate (n) at (1.9, 1.9);
			\coordinate (t) at (1.1, 2.2);
			\coordinate (b) at (0.7, 1.43);
			\draw[->,thin,gray] (origin)--(n) node[anchor=west]{};
			\draw[->,thin,gray] (origin)--(t) node[anchor=south]{};
			\draw[->,thin,gray] (origin)--(b) node[anchor=east]{};
			
			\node[] at (0.7, 0.6) (label) {b)};
			\end{tikzpicture}
			\begin{tikzpicture}[>={Stealth[inset=3pt,length=8pt,angle'=28,round]}]
			\node[anchor=south west,inner sep=0] at (0,0) {
				\includegraphics[trim=0.20\WSSurf{} 0.2\HSSurf{} 0.3\WSSurf{} 0.28\HSSurf{},clip,width=0.3\columnwidth] {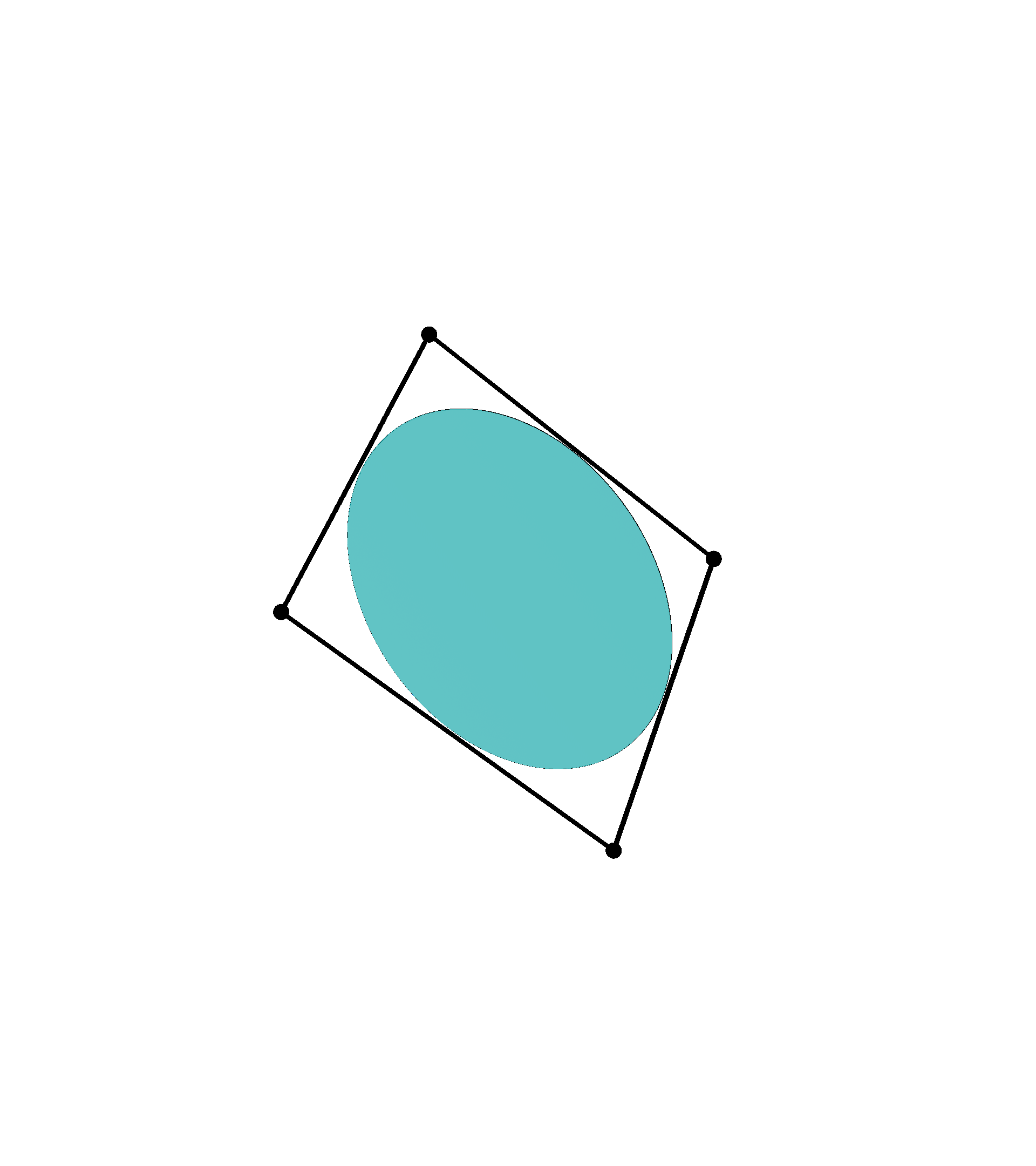}
			};
			\node[] at (0.7, 0.6) (label) {c)};
			\end{tikzpicture}
			
		}

		\caption{Surfel rendering is done in three steps. a) the vertex shader creates a basis from the normal, tangent and bitangent vectors. b) the geometry shader creates from each vertex a rectangle orientated according to the basis. c) the fragment shader creates the elliptical shape by discarding the fragments in the corners of the rectangle. }
		\label{fig:surfelCreate}
	\end{figure}
	\egroup

	 
	Once the surfels are created, they are rendered into the G-Buffer. Surfels that overlap within a small distance to each other accumulate their attributes and increment a weight for the current pixels that will be used later for normalization.
	
	During surfel rendering, the G-Buffer is changed from being stored as unsigned bytes to half floats in order to support the accumulation of attributes for overlapping surfels.
	
	The composing pass then normalizes the G-Buffer by dividing the accumulated albedo, normals, metalness and roughness by the weight stored in the alpha channel of the albedo. 
	
	Finally, composing proceeds as before with the PBR and \ac{IBL} pipeline. This yields similar results as mesh rendering which can be seen in~\reffig{fig:surfel}.

		\bgroup
	
	\definecolor{surf-cyan}{RGB}{58, 185, 201}
	\def\W{\columnwidth}
	\setlength{\tabcolsep}{1pt}
	\def\ImgBigMesh{./imgs/surfel/bunny_mesh.png}
	\def\ImgSmallMesh{./imgs/surfel/bunny_closeup_wireframe_3.jpg}
	\def\ImgBigSurfel{./imgs/surfel/bunny_surfel.png}
	\def\ImgSmallSurfel{./imgs/surfel/bunny_closeup_surfel_4.jpg}
	\newlength{\WSurf}
	\newlength{\HSurf}
	\settowidth{\WSurf}{\includegraphics{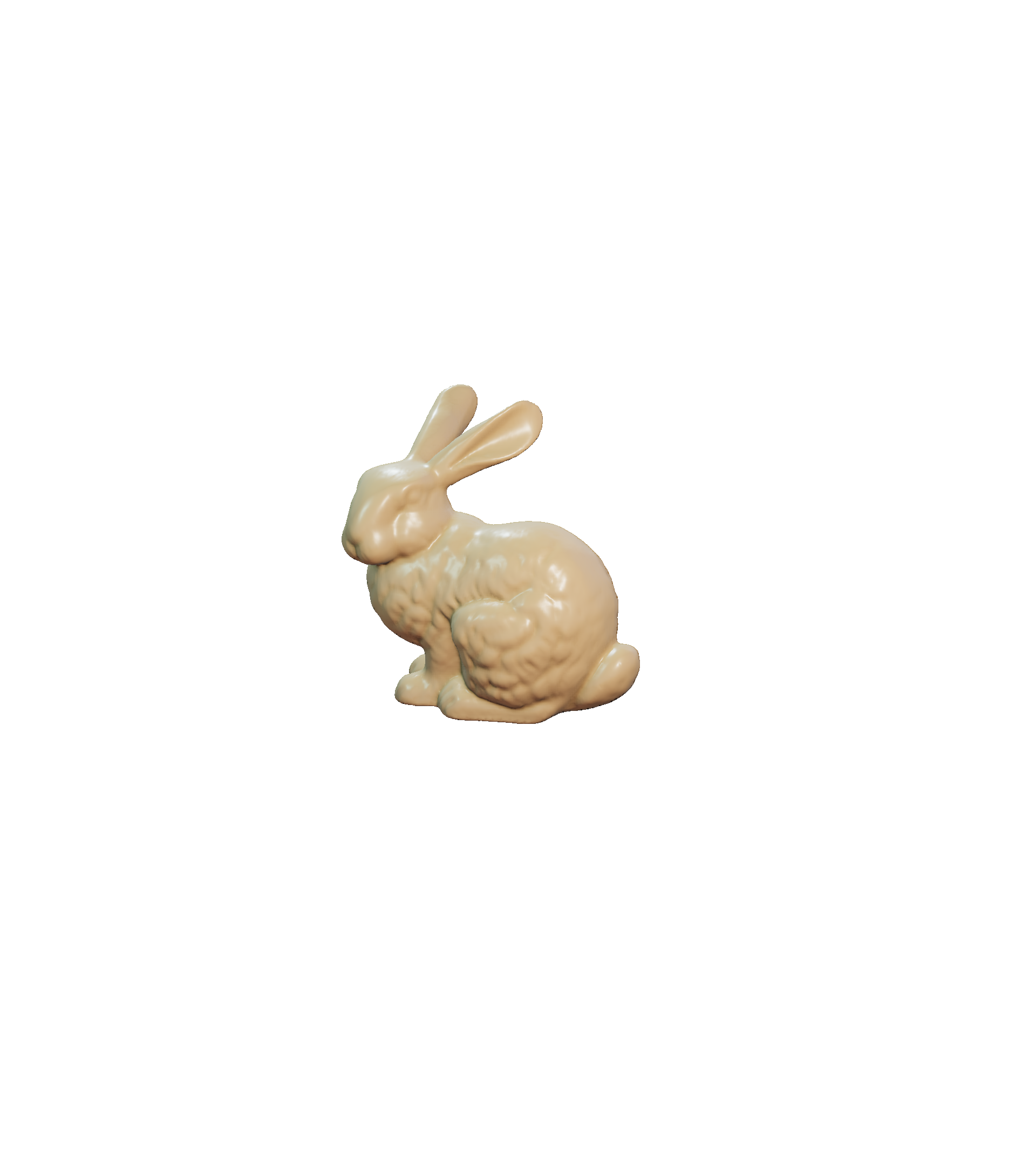}}
	\settoheight{\HSurf}{\includegraphics{\ImgBigMesh}}
	\begin{figure}
		\hskip-0.3cm 
		\resizebox{1.08\columnwidth}{!}{
			\begin{tabular}{C{\W} C{\W}}
				\centering
				\setlength{\tabcolsep}{1pt}
				\begin{tikzpicture}
				
				\begin{scope}[spy using outlines={circle, magnification=4, size=3.7cm, connect spies}][spy using outlines={rectangle,yellow,magnification=2,size=1.5cm}]
				
				\node at (0.0,0) {\includegraphics[align=c,trim=.33\WSurf{} .35\HSurf{} .33\WSurf{} .33\HSurf{},clip, width=.43\columnwidth]{\ImgBigMesh}} ;
				\spy [surf-cyan, every spy on node/.append style={line width=0.7mm}, every spy in node/.append style={line width=1.0mm} ] on (-0.4, -0.4) in node(spy) [left] at (4.9, 0.1);
				\end{scope}
				
				\node[circle, minimum size=3.6cm,
				path picture={
					\node at (path picture bounding box.center){
						\includegraphics[align=c, width=.475\columnwidth]{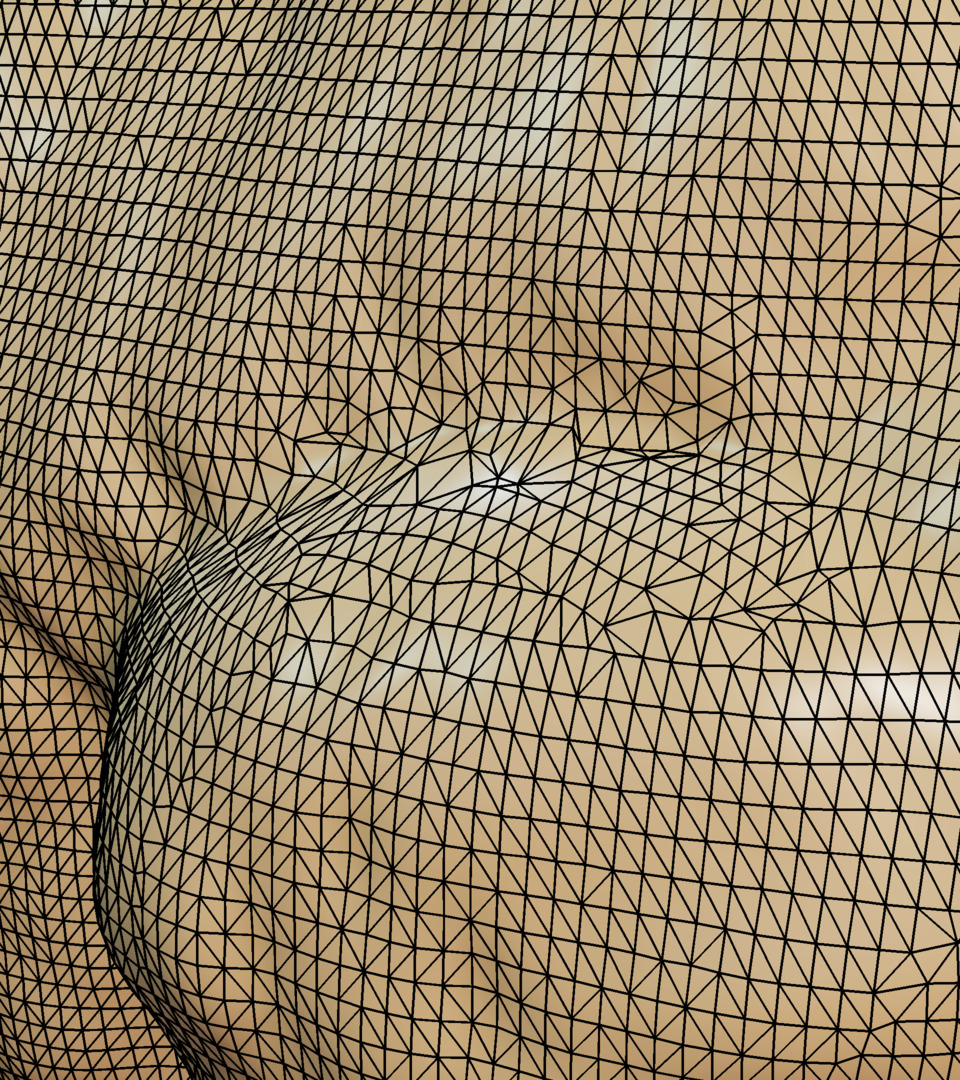}
					};}
				] (circle) at (spy.center) {						
				};
				\end{tikzpicture} &
				
				\begin{tikzpicture}
				
				\begin{scope}[spy using outlines={circle, magnification=4, size=3.7cm, connect spies}][spy using outlines={rectangle,yellow,magnification=2,size=1.5cm}]
				\node {\includegraphics[align=c,trim=.33\WSurf{} .35\HSurf{} .33\WSurf{} .33\HSurf{},clip, width=.43\columnwidth]{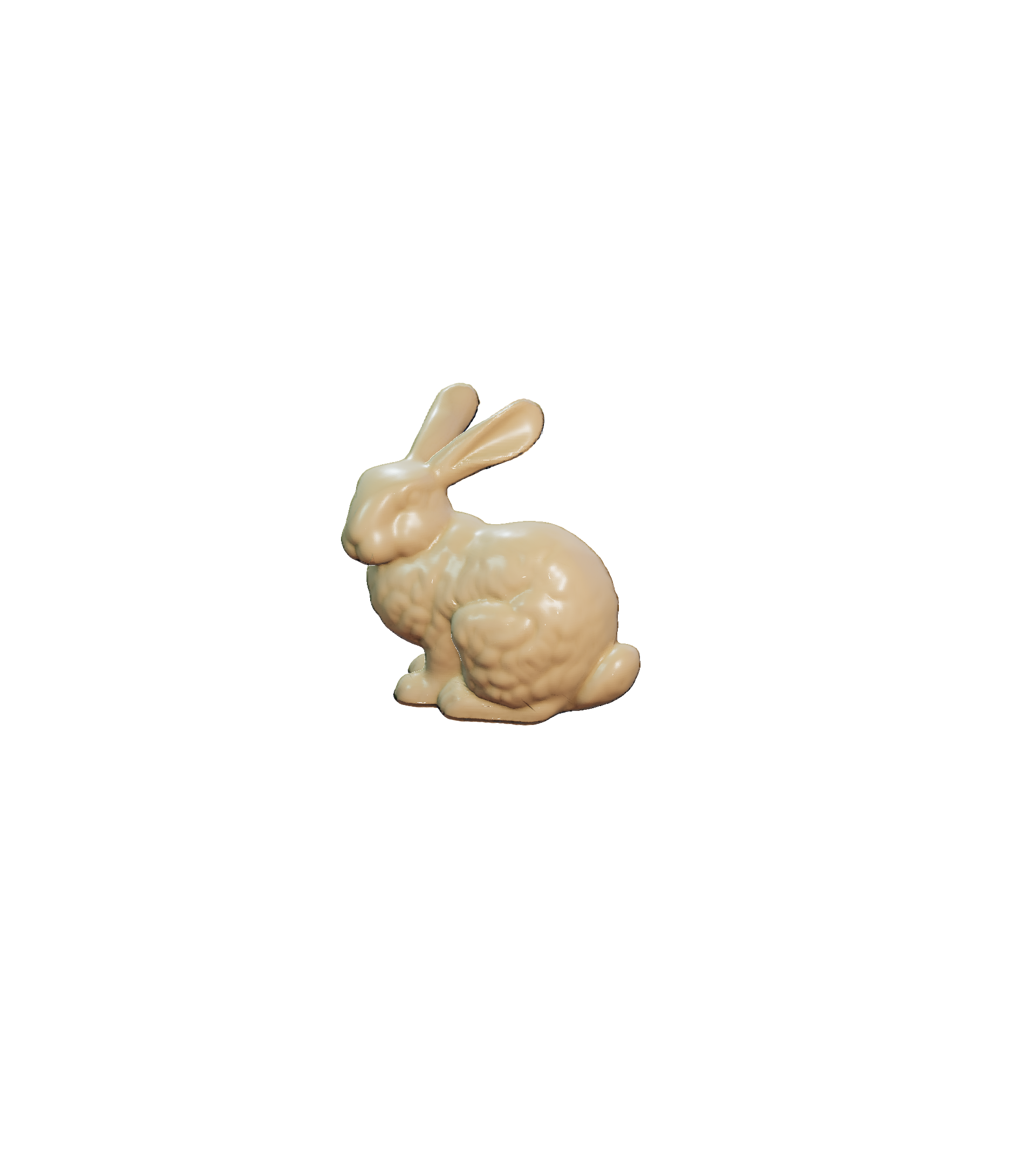}};
				\spy [surf-cyan, every spy on node/.append style={line width=0.7mm}, every spy in node/.append style={line width=1.0mm} ] on (-0.4, -0.4) in node(spy) [left] at (4.9, 0.1);
				\end{scope}
				
				\node[circle, minimum size=3.6cm,
				path picture={
					\node at (path picture bounding box.center){
						\includegraphics[align=c, width=.475\columnwidth]{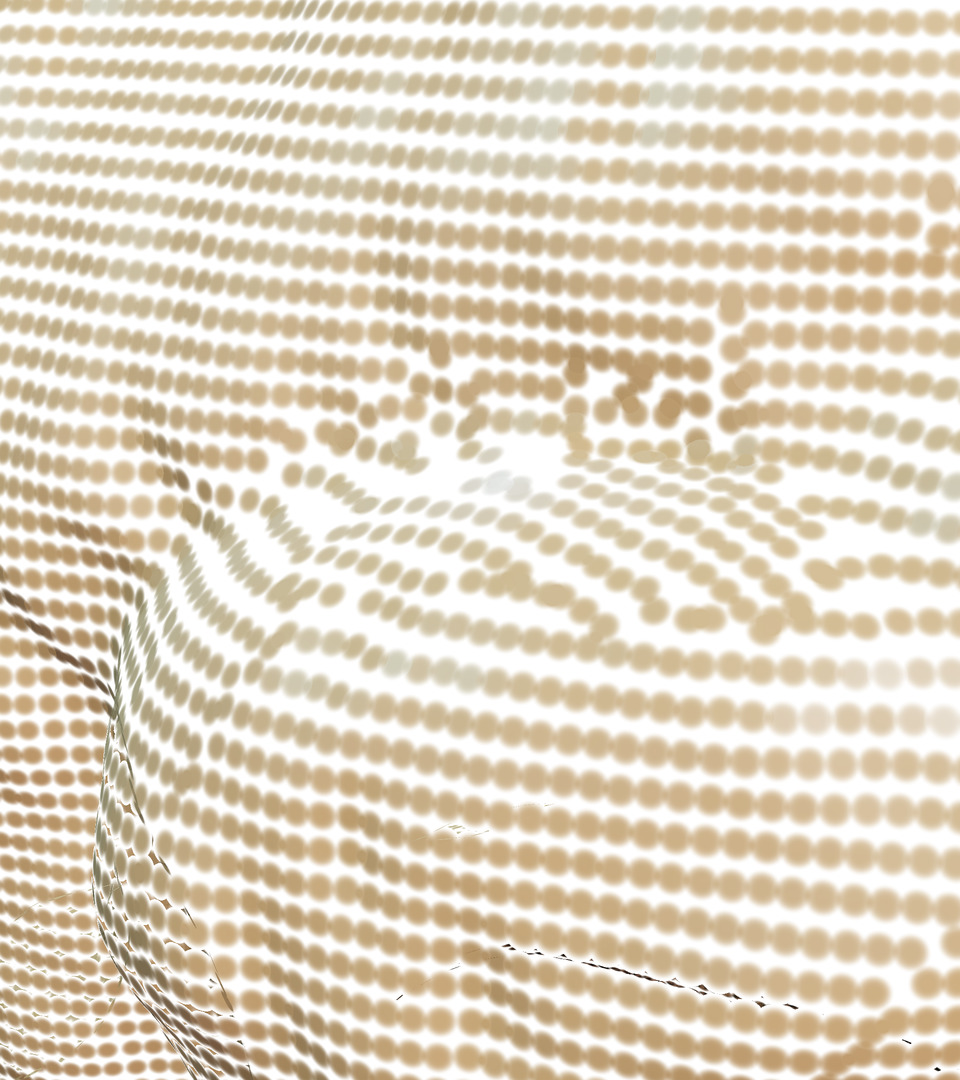}
					};}
				] (circle) at (spy.center) {						
				};
				\end{tikzpicture}\\

			\end{tabular}
		}
		\caption{Comparison between mesh and surfel rendering. For clarity, we reduce the radius of the surfels in the zoomed-in view.}
		\label{fig:surfel}
	\end{figure}
	\egroup

	\subsection{Line Rendering}

	Line rendering is useful for showing the wireframe of meshes or for displaying edges between arbitrary vertices indicated by the $\mathbf{E}$ matrix. 
	We perform line rendering by forward rendering directly into the final image as we do not want lines to be affected by lighting and shadowing effects.

\section{Effects}
	Multiple post-processing effects that are supported in EasyPBR: shadows, \ac{SSAO}, bloom.

	\subsection{Shadows}
		EasyPBR supports point-lights, which can cast realistic soft shadows onto the scene. Shadows computation is performed through shadow mapping~\cite{williams1978casting}.
		The process works by first rendering the scene only as a depth map into each point-light as if they were a normal camera.
		
		Afterwards, during compositing, we check if a fragment's depth is greater than the depth recorded by a certain light. If it is greater, then the fragment lies behind the surface lit by the light and is therefore in shadow. 
		In order to render soft shadows, we perform Percentage Closer Filtering~\cite{reeves1987rendering} by checking not only the depth around the current fragment but also the neighboring ones in a $3\times3$ patch in order to obtain a proportion of how much of the surface is shadowed instead of just a binary value. 
		
		The shadow maps are updated only if the objects in the scene or the lights move. While the scene remains static, we use the last rendered depth map for each light. This constitutes a significant speed-up in contrast to the naive approach of recomputing the shadow map at every frame.
		
		Shadows from multiple lights interacting with the scene can be observed in~\reffig{fig:effects_shadow}.
		
	\subsection{Screen-space Ambient Occlusion}	
		Ambient occlusion is used to simulate the shadowing effect caused by objects blocking the ambient light. Simulating occlusion requires global information of the scene geometry and is usually performed through ray-tracing, which is costly to compute. Screen-space ambient occlusion addresses this issue by using only the current depth buffer as an approximation of the scene geometry, therefore avoiding the use of costly global information and making the ambient occlusion real-time capable. The effect of \ac{SSAO} can be viewed in~\reffig{fig:effects_shadow}. 
		
		Our \ac{SSAO} implementation is based on the Normal-oriented Hemisphere method~\cite{bavoil2008screen}. After creating the G-Buffer, we run the \ac{SSAO} pass in which we randomly take samples along the hemisphere placed at each pixel location and orientated according to the normal stored in the buffer. The samples are compared with the depth buffer in order to get a proxy of how much the surface is occluded by neighboring geometry. The \ac{SSAO} effect is computed at half the resolution of the G-Buffer and bilaterally blurred in order to remove high-frequency noise caused by the low sample count. 
	
		\bgroup
		\def\ImgShad{./imgs/effects/shadows_only_2.png} 
		\newlength{\WSShad}
		\newlength{\HSShad}
		\settowidth{\WSShad}{\includegraphics{\ImgShad}}
		\settoheight{\HSShad}{\includegraphics{\ImgShad}}
		\def\ImgBloom{./imgs/effects/bloom_only_2_bg.png} 
		\newlength{\WSBloom}
		\newlength{\HSBloom}
		\settowidth{\WSBloom}{\includegraphics{\ImgBloom}}
		\settoheight{\HSBloom}{\includegraphics{\ImgBloom}}
		\begin{figure}[]
			\captionsetup[subfloat]{labelformat=empty}
			\centering
			
			\subfloat[a) Shadows and SSAO ]{
				\includegraphics[trim=0.25\WSShad{} 0.13\HSShad{} 0.25\WSShad{} 0.14\HSShad{},clip,width=0.5\columnwidth] {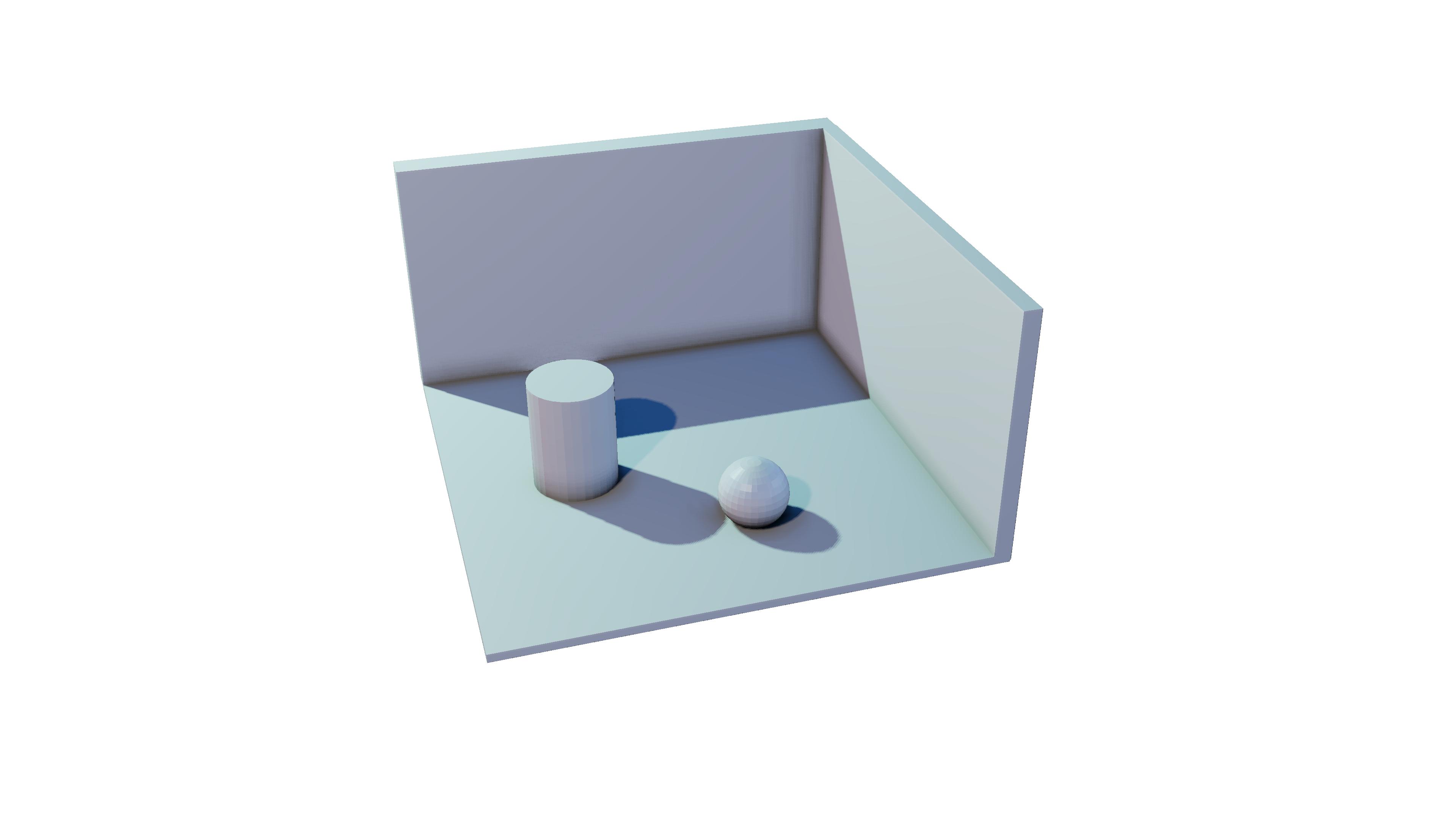}
				\label{fig:effects_shadow}
			}
			\subfloat[b) Bloom]{
				\includegraphics[trim=0.26\WSBloom{} 0.26\HSBloom{} 0.26\WSBloom{} 0.23\HSBloom{},clip,width=0.37\columnwidth] {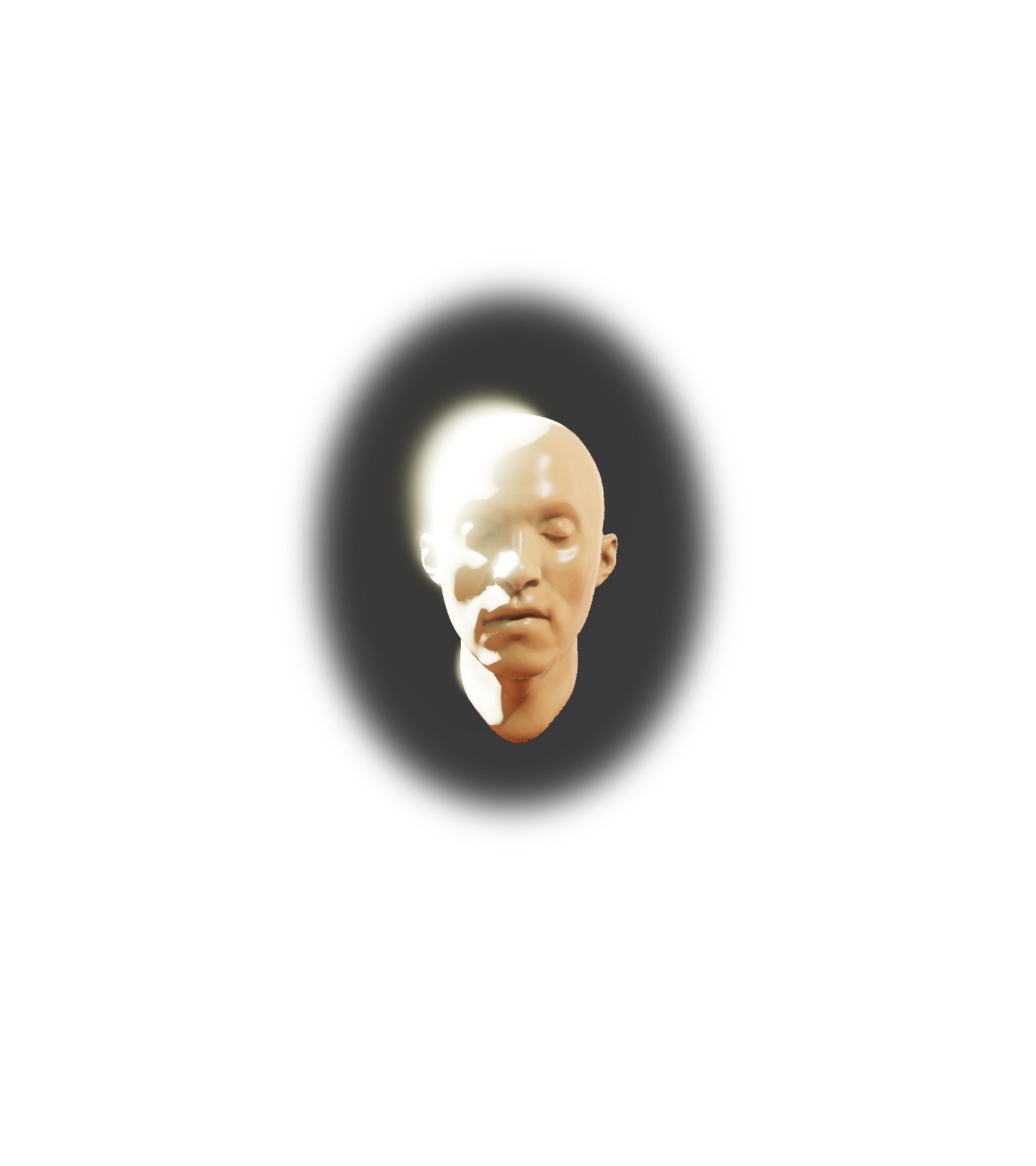}
				\label{fig:effects_bloom}
			}
	
			\caption{ Various post-processing effects can be enabled in the renderer. Soft shadows and ambient occlusion convey a sense of depth, and bloom simulates the light bleed from bright parts of the scene like the sun or reflective surfaces.}
			\label{fig:effects}
		\end{figure}
		\egroup
	
	\subsection{Bloom}
		Bloom is the process by which bright areas of the images bleed their color onto adjacent pixels. This can be observed, for example, with very bright sunlight, which causes the nearby parts of the image to increase in brightness. 
		Bloom is implemented by rendering into a bright-map only the parts of the scene that are above a certain level of brightness. 
		
		This bright map would now need to be blurred with a Gaussian kernel and then added on top of the original image. However, performing blurring at the resolution of the full screen is too expensive for real-time purposes, and we, therefore, rely on approximations. We create an image pyramid with up to six levels from the bright map. We blur each pyramid level starting from the second one upwards. 
		Blurring by using an image pyramid allows us to use very large kernels.
		
		Finally, the bright-map pyramid is added on top of the original image in order to create a halo-like effect. The result can be seen in~\reffig{fig:effects_bloom}.
		 
	\subsection{Final Compositing}
		The compositing is the final rendering pass before showing the image to the screen. It takes all the previous rendering passes (G-Buffer, \ac{SSAO}, etc.) and combines them to create the final image. 
		Finally, after creating the composed image, it needs to be tone-mapped and gamma-corrected in order to bring the \ac{HDR} values into a \ac{LDR} range displayable on the screen. For this, we use the \ac{ACES} tone mapper due to its high-quality filmic look. We further offer support for the Reinhard~\cite{reinhard2002photographic} tone mapper.

\section{Automatic parameters} 
	Various parameters govern the rendering process. The user can leave them untouched, and our rendering tool will try to make an educated guess for them at runtime. 

	By default, EasyPBR creates a 3-point light setup consisting of a key light that provides most of the light for the scene, a fill light softening the shadows, and a rim light placed behind the object to separate it from the background. 
	The distances from the object center towards the light are determined such that the scene radiance at the object has a predefined value. This makes the lighting setup agnostic to scaling of the mesh, so EasyPBR can out of the box render any kind of mesh regardless of the unit system it uses. At any point at runtime, the user can tweak the position, intensity, and color of the lights. 
	
	The camera is placed in the world so that the entire object is in view. Also, the near and far planes of the cameras are set according to the scale of the scene.
	
	\ac{SSAO} radius is a function of the scene scale. By default, we choose the radius to be $5\%$ of the scale of the scene to be rendered.
	
	The rendering mode depends on the object in the scene. If the object has no connectivity provided as triangles in the $\mathbf{F}$ matrix, then we render it as a point cloud using \ac{EDL}. Otherwise, we render it as a mesh. If the user provides normals and tangent vectors, then we render it as a series of surfels. This ensures that whatever data we put in, our objects will be visualized in an appropriate manner.

\section{Performance} 

	We evaluate the performance of EasyPBR and compare it to Meshlab and VTK, as they are common tools used for visualization. We run all three tools on an Nvidia RTX 2060. As a metric, we use the milliseconds per frame and test with two meshes, one high-resolution mesh with 23 million faces (Goliath statue from~\reffig{fig:recordCam}) and the 3D scanned head (\reffig{fig:heads}) with half a million faces and high-resolution 8K textures. The results are shown in~\reftab{tab:perf}.

	First, we remark that Meshlab v1.3.2, the version that is available in the Ubuntu 18.04 repositories, struggles to render the Goliath mesh, requiring almost $500\si{\milli\second}$. This is due to an internal limitation on the amount of memory that is allowed for the geometry. Once the mesh uses more memory than this internal threshold, Meshlab silently switches to immediate mode rendering, which causes a significant performance drop. Newer versions of Meshlab ( version 2020.09 ) have to be compiled from source, but they allow to increase this memory threshold above the default $350\si{\mega\byte}$ and render the mesh at $6\si{\milli\second}$ per frame.

	We point out that Meshlab is faster than both approaches due to the usage of only simple Phong shading.

\begin{table}
\vspace*{5pt} 
	\caption{Timings in milliseconds to render a frame.} \label{tab:perf}
\begin{center}
	\footnotesize
	\begin{tabular}{ c | c   c  c  c}
		 		& EasyPBR & VTK & \shortstack{Meshlab \\ \scriptsize{v2020.09}} & \shortstack{Meshlab \\ \scriptsize{v1.3.2}}\\  
		 \toprule
		Goliath & 6.2 & 6.1  & 6.0 & 558\\  
		\midrule
		Head & 1.6 & 1.6  & 1.1 & 1.1\\
		\bottomrule
	\end{tabular}
\end{center}
\end{table}

\section{Applications}
	
	The flexibility offered by EasyPBR allows it to be used for a multitude of applications. We gather here a set of real cases in which it was used.

	\subsection{Synthetic Data Generation}
		Deep learning approaches require large datasets in order to perform supervised learning, and the effort in annotating and labeling such datasets is significant. Consequently, interest has recently increased in using synthetic data to train the models and thus avoid or reduce the need for real labeled data. 

		EasyPBR has been used in the context of deep learning to create realistic 2D images for object detection tasks. Specifically, it has been used for training a drone detector capable of recognizing a drone in mid-flight. The model requires large amounts of data in order to cope with the variations in lighting, environment conditions, and drone shape. EasyPBR was used to create realistic environments in which we placed various drone types that were rendered together with ground truth bounding boxes annotations.

		An example of a synthetic image and the output from the drone detector model can be seen in~\reffig{fig:synth}.
		
		\bgroup
		\def\ImgDet{./imgs/synthetic/copter_ball_detections_02.jpg} 
		\newlength{\WDet}
		\newlength{\HDet}
		\settowidth{\WDet}{\includegraphics{\ImgDet}}
		\settoheight{\HDet}{\includegraphics{\ImgDet}}
		\begin{figure}[]
			\captionsetup[subfloat]{labelformat=empty}
			\centering
			
			\subfloat[a) Synthetic DJI M100 drone]{
				\includegraphics[width=0.47\columnwidth] {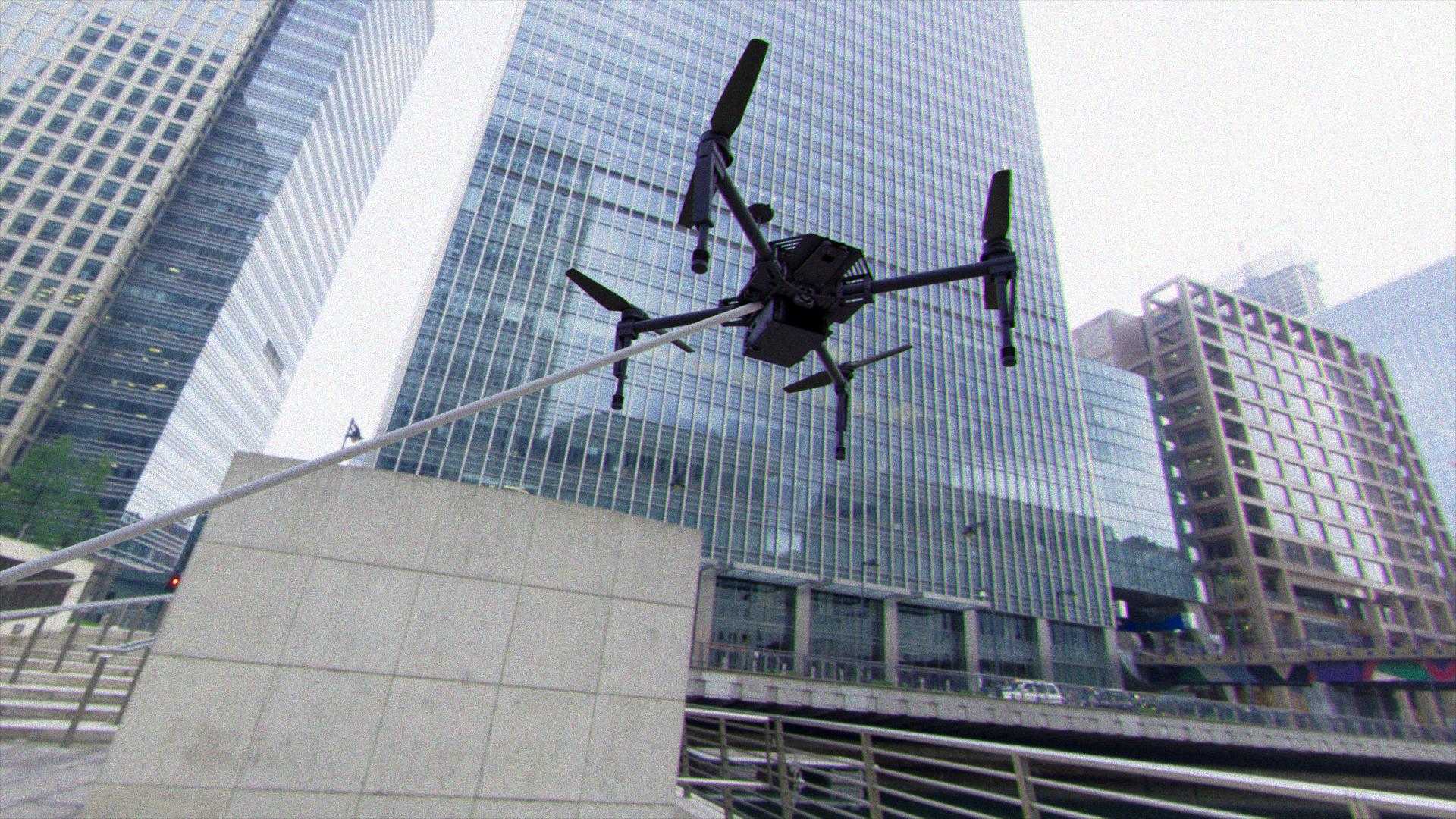}
			}
			\subfloat[b) Detection in real image ]{
				\includegraphics[trim=0.3\WDet{} 0.4\HDet{} 0.3\WDet{} 0.2\HDet{},clip,width=0.47\columnwidth] {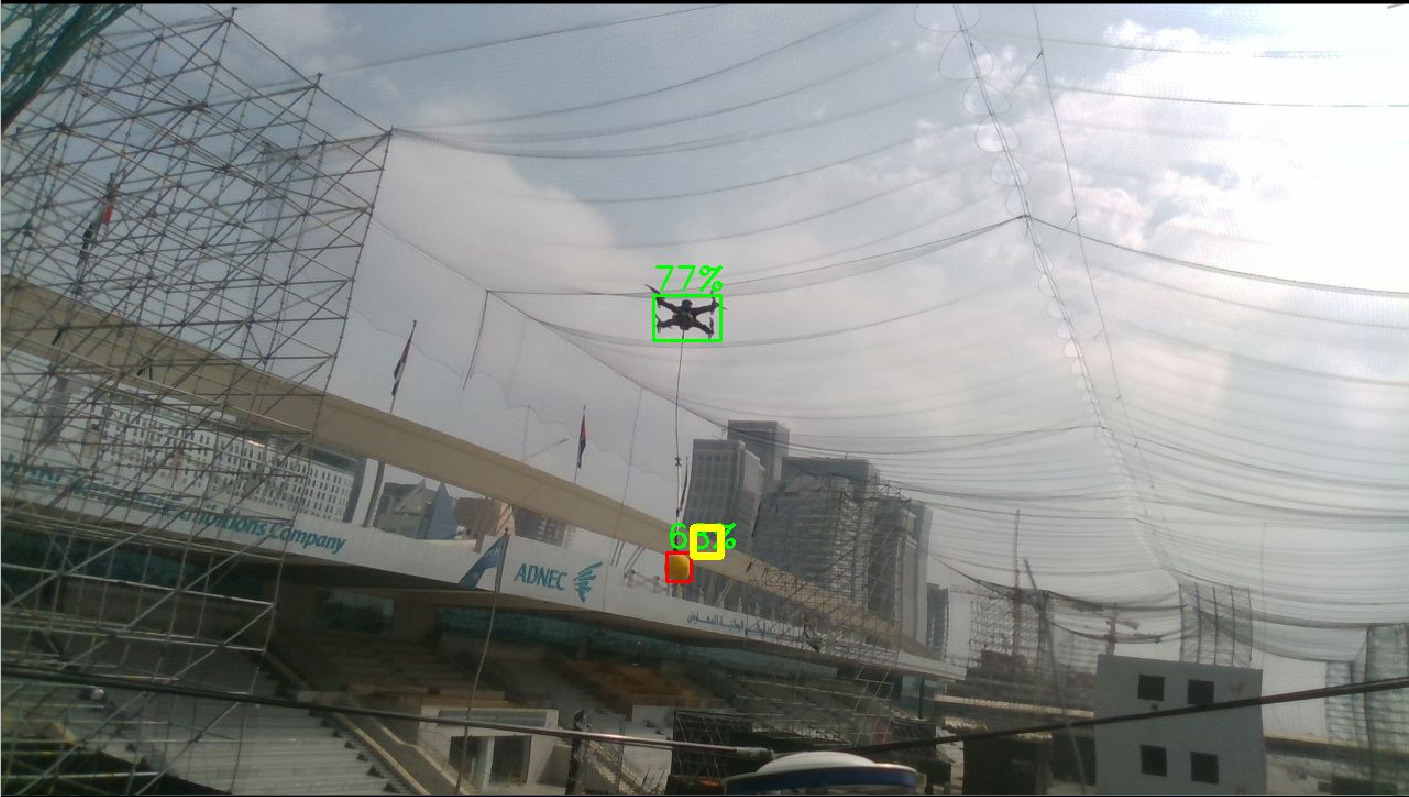}
			}
			
			\caption{ Synthetic data can be easily rendered and used for deep learning applications. Images of drones together with ground truth bounding box annotation were rendered and used for training a drone detector.
			}
			\label{fig:synth}
		\end{figure}
		\egroup 
		
		The core of the Python code used to render the synthetic images can be compactly expressed as:
		\vbox{ 
		\begin{lstlisting}[style=easypbr,xleftmargin=-.01\columnwidth]
	view  = Viewer()
	view.load_environment_map("./map.hdr")
	drone = Mesh("./drone.ply")
	Scene.show(drone, "drone")
	view.recorder.record("img.png")
		\end{lstlisting}
		}
%
%
	\subsection{Visualizer for 3D Deep Learning}

		Many recent 3D deep learning applications take as input either raw point clouds or voxelized clouds. Visually inspecting the inputs and outputs of the network is critical for training such models. 
		
		EasyPBR interfaces with PyTorch~\cite{paszke2017automatic} and allows for conversion between the CPU data of the point cloud and GPU tensors for model input and output. EasyPBR is used for data loading by defining a parallel thread that reads point cloud data onto the CPU and then uploads to GPU tensors. After the model processes the tensors, the prediction is directly read by EasyPBR and used for visualization. An example of 3D semantic segmentation and instance segmentation of point clouds can be seen in~\reffig{fig:lattice} and~\reffig{fig:plants} where our tool was used for visualization and data loading.

		\bgroup
		\def\ImgSem{./imgs/semantickitti/247.jpg} 
		\newlength{\WSSem}
		\newlength{\HSSem}
		\settowidth{\WSSem}{\includegraphics{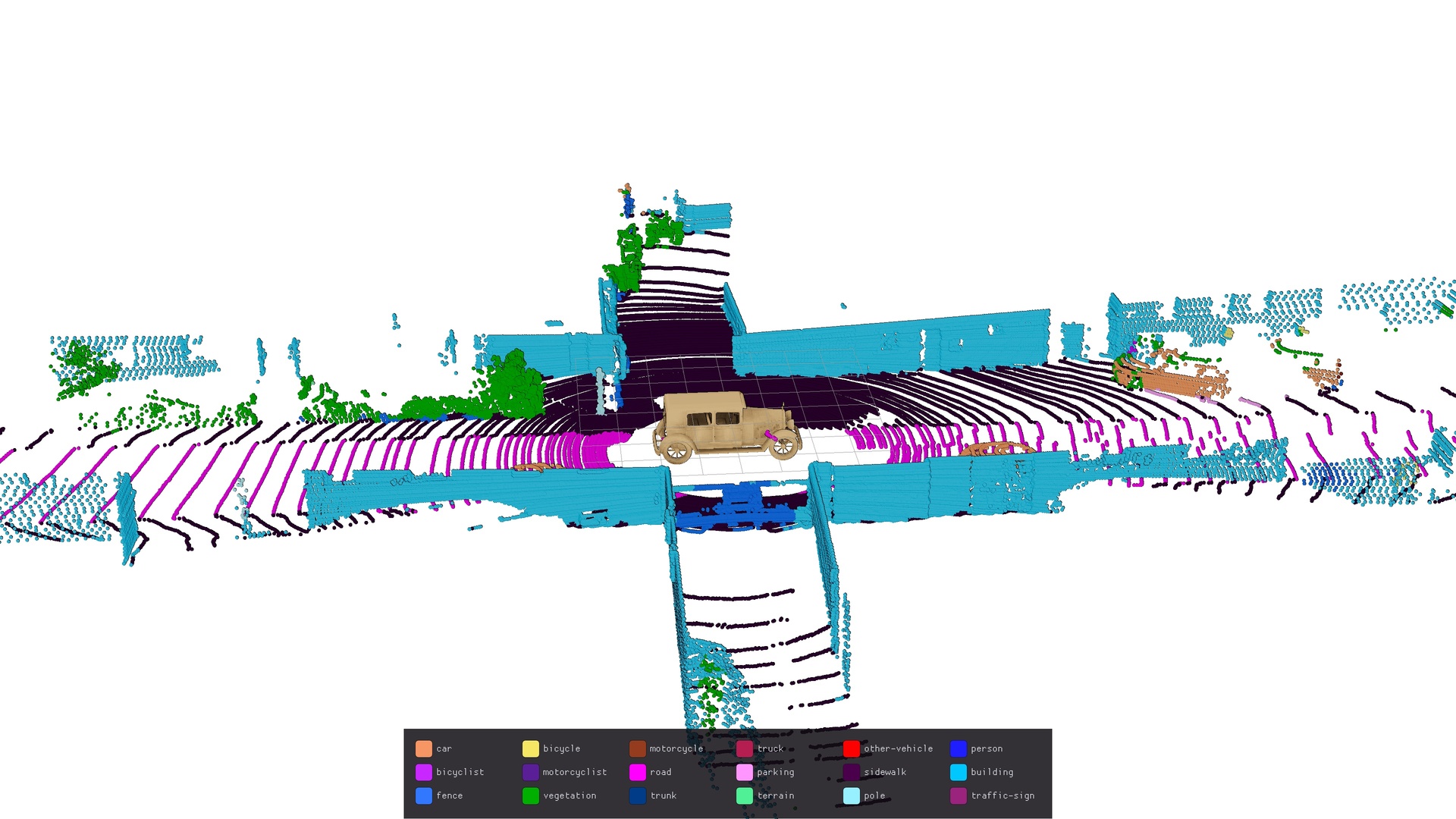}}
		\settoheight{\HSSem}{\includegraphics{\ImgSem}}
		\begin{figure}[]
			\centering
			\includegraphics[angle=-2, trim=0.1\WSSem{} 0.3\HSSem{} 0.15\WSSem{} 0.34\HSSem{},clip,width=0.9\columnwidth] {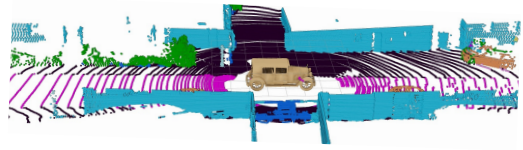}
			
			\caption{ Point cloud segmented by LatticeNet~\cite{rosu2019latticenet} and visualized with the colormap of SemanticKITTI~\cite{behley2019semantickitti}. }
			\label{fig:lattice}
		\end{figure}
		\egroup
		
		\bgroup
		\def\ImgPlant{./imgs/plants/phen_instance_cloud.png} 
		\newlength{\WP}
		\newlength{\HP}
		\settowidth{\WP}{\includegraphics{\ImgPlant}}
		\settoheight{\HP}{\includegraphics{\ImgPlant}}
		\begin{figure}[]
			\captionsetup[subfloat]{labelformat=empty}
			\centering

			\begin{minipage}[c]{0.3\columnwidth}
				\includegraphics[trim=0.35\WP{} 0.30\HP{} 0.4\WP{} 0.3\HP{},clip,width=1.0\columnwidth] {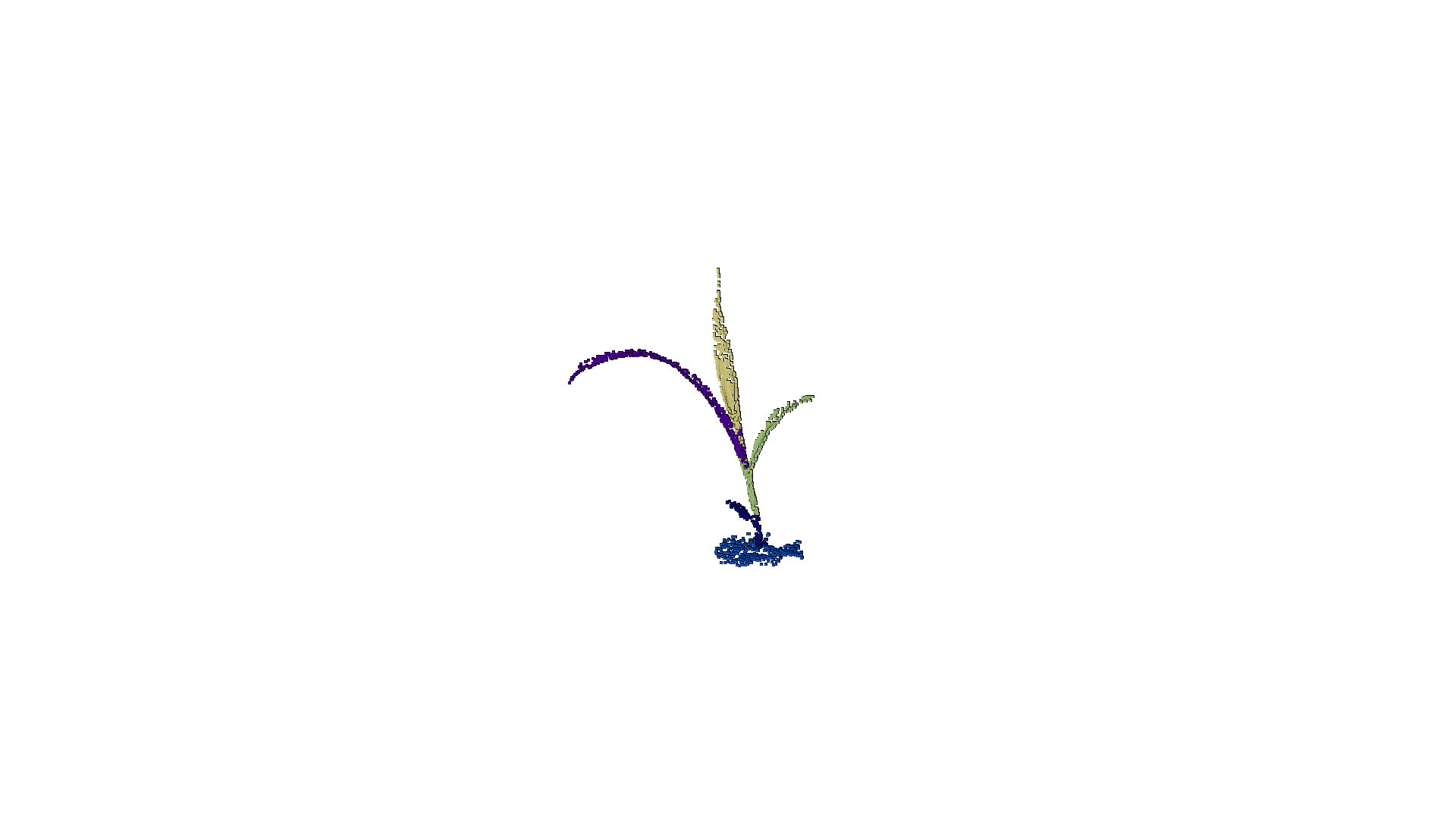}
			\end{minipage}
			\begin{minipage}[c]{0.3\columnwidth}
				\includegraphics[trim=0.35\WP{} 0.33\HP{} 0.35\WP{} 0.25\HP{},clip,width=1.0\columnwidth] {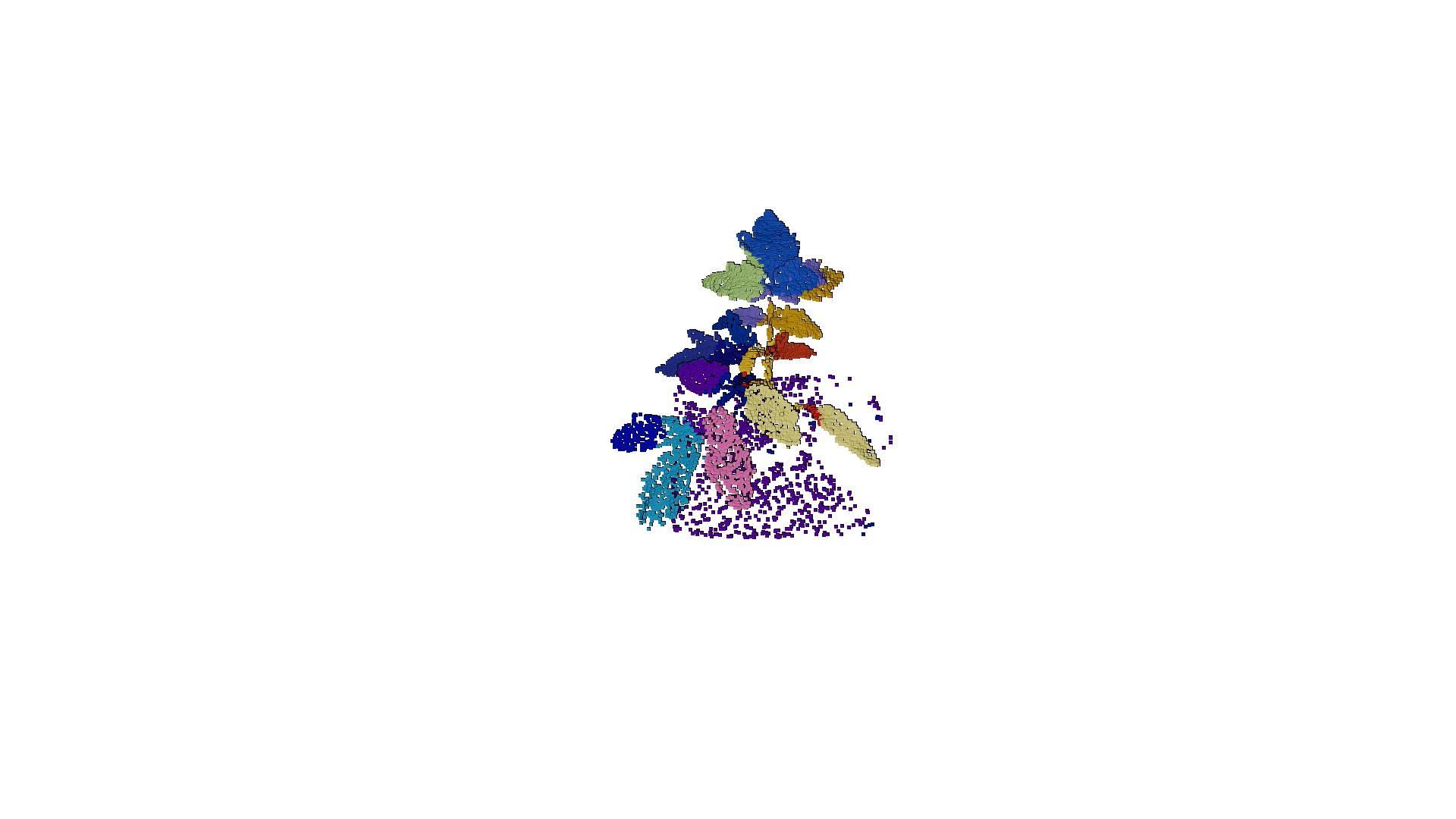} 
			\end{minipage}
			\begin{minipage}[c]{0.3\columnwidth}
				\includegraphics[trim=0.45\WP{} 0.3\HP{} 0.4\WP{} 0.45\HP{},clip,width=1.0\columnwidth] {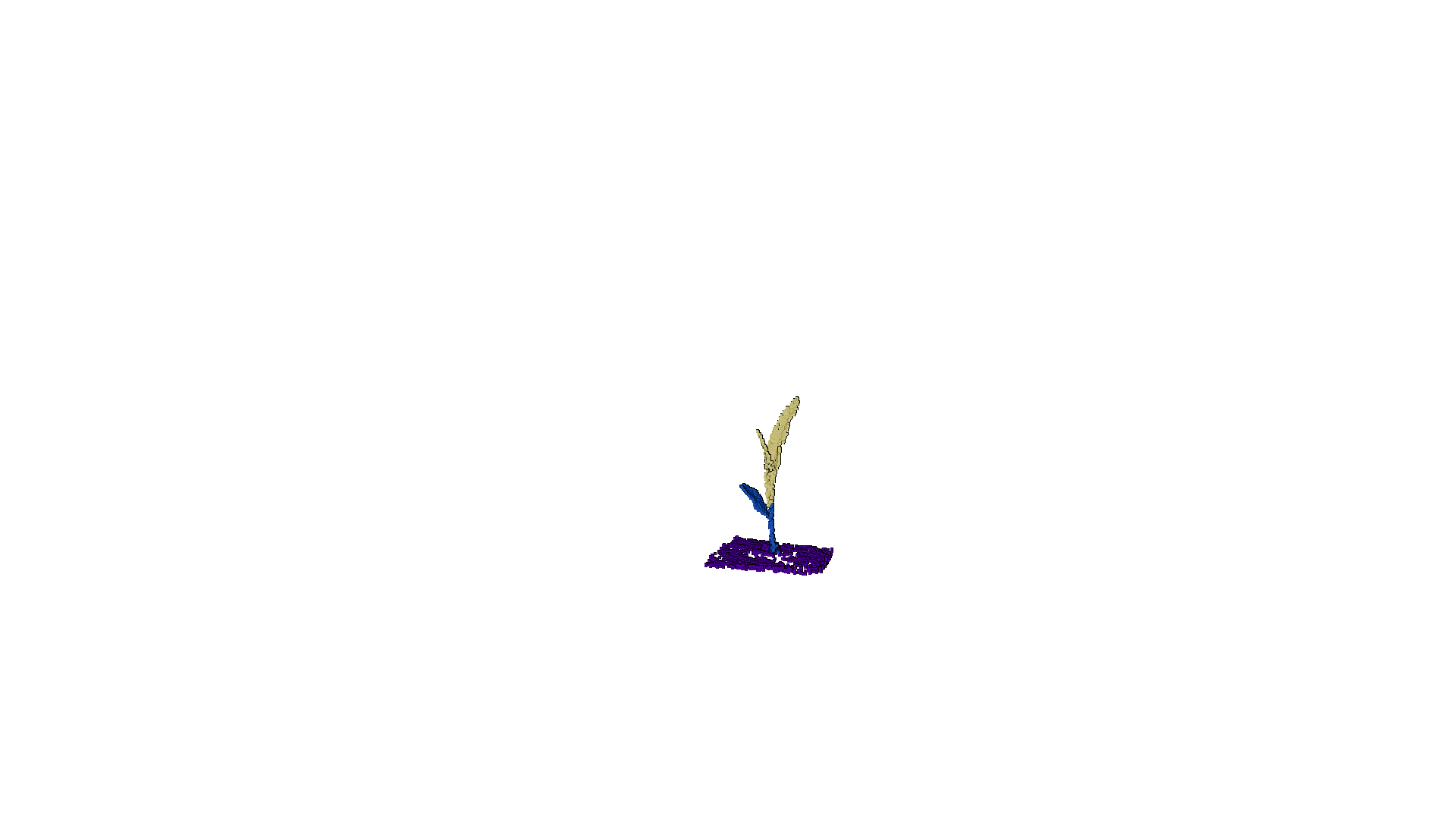} 
			\end{minipage}

			\caption{ Instance segmentation of plant leaves using LatticeNet~\cite{rosu2019latticenet}.
			}
			\label{fig:plants}
		\end{figure}
		\egroup 
		
		Inside the training loop of a 3D deep learning approach, Python code similar to this one can be used for visualization and input to the network: 
		\vbox{
		\begin{lstlisting}[style=easypbr,xleftmargin=.05\columnwidth]
	cloud   = Mesh("./lantern.obj")
	points  = eigen2tensor(cloud.V)
	pred    = net(points)
	cloud.L = tensor2eigen(pred)
	Scene.show(cloud, "cloud")
		\end{lstlisting}
		}

	\subsection{Animations} 
		EasyPBR can also be used to create simple 2D and 3D animations. The 3D viewer keeps a timer, which starts along with the creation of the application. At any point, the user can query the delta time since the last frame and perform incremental transformations on the objects in the scene.

		Additionally, the user can create small rigid kinematic chains by specifying a parent-child hierarchy between the objects. Transformations of the parent object will, therefore, also cause a transformation of the child. This is useful when an object is part of another one. 

%
	\subsection{Recording} 
		EasyPBR can be used both for taking screenshots of the scene and for recording movies while the virtual camera is moving through the environment. Through the GUI, the user can place a series of key-poses through which the camera should move. The user then specifies the time to transition from one pose to another and lets the animation run. The camera linearly interpolates between the specified $\mathbf{SE}(3)$ poses while continuously recording. The images saved can then be converted into a movie.
		
		An example of the camera trajectory surrounding an object to be captured can be seen in~\reffig{fig:recordCam}.
		
		\bgroup
		\begin{figure}[]
			\centering
			\includegraphics[width=0.8\columnwidth] {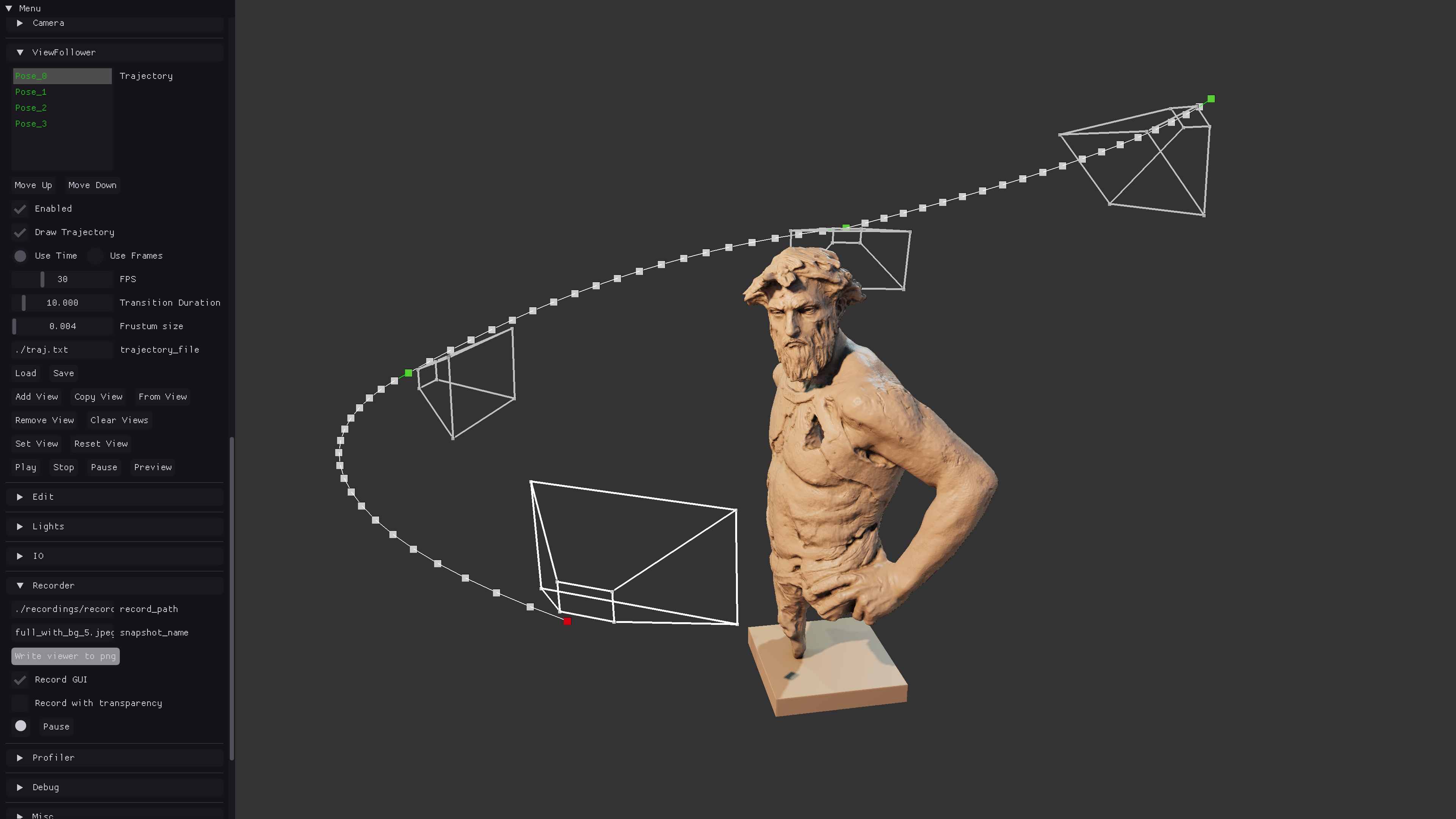}
			
			\caption{ Viewer GUI and camera trajectory for recording a video of the 3D object. }
			\label{fig:recordCam}
		\end{figure}
		\egroup

\section{Conclusion}

We presented EasyPBR, a physically-based renderer with a focus on usability without compromising visual quality. Various state-of-the-art rendering methods were implemented and integrated into a framework that allows easy configuration. EasyPBR simplifies the rendering process by automatically choosing parameters to render a specific scene, alleviating the burden on the user side. 

In future work, we intend to make EasyPBR easier to integrate for remote visualizations and also add further effects like depth of field and transparencies.

We make the code fully available together with the scripts to create all the figures shown in this paper. We hope that this tool will empower users to create visually appealing and realistic images without sacrificing performance or enforcing the burden of a steep learning curve.

\bibliographystyle{apalike}
\interlinepenalty=10000 
{\small
	\bibliography{references}}

\end{document}